\documentclass[useAMS,usenatbib]{mn2e}

\usepackage{amsmath}
\usepackage{amssymb}
\usepackage{graphicx}
\usepackage{graphics}
\usepackage{pdflscape}
\usepackage{float}
\usepackage{epstopdf}
\usepackage{multicol}
\usepackage{breakurl}
\usepackage{textcomp}
\usepackage{enumitem}
\usepackage{subfigure}
\usepackage[bf,labelsep=period,font=small]{caption}
\usepackage{subfloat}
\usepackage{subfigure}
\usepackage{gensymb}
\usepackage{supertabular}
\usepackage{longtable,ltxtable}
\usepackage{afterpage}
\usepackage[flushleft]{threeparttable}
\title[Spectroscopic survey of \emph{Kepler} A stars]{A spectroscopic test of the rotational modulation origin of periodic \emph{Kepler} photometric variability of A-type stars}
\author[J. Sikora et al.]
	{J.~Sikora,$^{1,2,3}$\thanks{Based on observations obtained at the Canada-France-Hawaii Telescope (CFHT) which is operated by the National Research Council of Canada, the Institut National des Sciences de l'Univers of the Centre National de la Recherche Scientifique of France, and the University of Hawaii.} G.~A.~Wade,$^{2}$ J.~Rowe,$^3$\\
$^{1}$Department of Physics, Engineering Physics \& Astronomy, Queen's University, Kingston, ON Canada, K7L 3N6\\
$^{2}$Department of Physics and Space Science, Royal Military College of Canada, PO Box 17000 Kingston, Ontario, Canada, K7K 7B4\\
$^{3}$Department of Physics and Astronomy, Bishop's University, Sherbrooke, Qu{\'e}bec, Canada, J1M 1Z7\\}

\begin{document}

\date{Submitted 2020 Mth. XX}

\pagerange{\pageref{firstpage}--\pageref{lastpage}} \pubyear{2020}

\maketitle

\label{firstpage}

\begin{abstract}
High-precision space-based photometry obtained by the \emph{Kepler} and \emph{TESS} missions has revealed evidence of rotational modulation associated with main sequence (MS) A and late-B type stars. Generally, such variability in these objects is attributed to inhomogeneous surface structures (e.g. chemical spots), which are typically linked to strong magnetic fields ($B\gtrsim100\,{\rm G}$) visible at the surface. It has been reported that $\approx44$~per~cent of all A-type stars observed during the \emph{Kepler} mission exhibit rotationally modulated light curves. This is surprising considering that $\lesssim10$~per~cent of all MS A-type stars are known to be strongly magnetic (i.e. they are Ap/Bp stars). We present a spectroscopic monitoring survey of 44 A and late-B type stars reported to exhibit rotational modulation in their \emph{Kepler} light curves. The primary goal of this survey is to test the hypothesis that the variability is rotational modulation by comparing each star's rotational broadening ($v\sin{i}$) with the equatorial velocities ($v_{\rm eq}$) inferred from the photometric periods. We searched for chemical peculiarities and binary companions in order to provide insight into the origin of the apparent rotational modulation. We find that 14 stars in our sample have $v\sin{i}>v_{\rm eq}$ and/or have low-mass companions that may contribute to or be responsible for the observed variability. Our results suggest that more than $10$~per~cent of all MS A and late-B type stars may exhibit inhomogeneous surface structures; however, the incidence rate is likely $\lesssim30$~per~cent.
\end{abstract}


\section{Introduction}\label{sect:intro}

Magnetic chemically peculiar stars (mCP or Ap/Bp stars) are known to exhibit horizontally inhomogeneous distributions of various chemical elements within their atmospheres \citep[e.g.][]{khokhlova1975,kochukhov2002a,silvester2014}. The formation of these intense, long-lived chemical abundance non-uniformities (i.e. ``chemical spots") can largely be attributed to the presence of strong, organized magnetic fields that are visible at the star's surface \citep{michaud1970,michaud1981}. One consequence of such chemical spots is that the star's flux is locally redistributed from the UV to longer wavelengths through line blanketing and backwarming \citep{kochukhov2005a}; when coupled with the star's rotation, chemical spots often lead to periodic photometric variability \citep[e.g.][]{krticka2009,krticka2012,krticka2015}.

Currently, only a relatively small subsample of main sequence (MS) A- and late B-type stars are known to exhibit low-frequency photometric variability that can convincingly be attributed to chemical spots. This subsample is dominated by the known mCP stars \citep[e.g.][]{adelman2007,bernhard2015}, which account for $\sim10$~per~cent of all MS A-type stars \citep{wolff1968,smith1971,sikora2019}. Therefore, the detection of rotational modulation associated with a MS A/late-B star generally implies a high probability that the star is magnetic \citep{buysschaert2018,david-uraz2019,david-uraz2020}. Based on an analysis of the high-precision light curves yielded by the \emph{Kepler} mission, \citet{balona2013} reported the surprising discovery that $\approx44$~per~cent of A-type stars in the \emph{Kepler} sample exhibit photometric variability that is consistent with a rotational modulation origin. These findings suggest that the accepted fraction of MS A-type stars that host detectable magnetic fields may be underestimated by $\approx400$~per~cent. Such a dramatic underestimation is made plausible by the fact that the reported photometric amplitudes, which have a mean value of $0.06\,{\rm mmag}$, are significantly lower than the detection limits associated with previous all-sky time series photometry surveys \citep[e.g. typical \emph{Hipparcos} photometric measurement uncertainties are $\gtrsim1\,{\rm mmag}$,][]{esa1997}.

Since 2009, four MS A-type stars -- Vega, Sirius~A, $\beta$ UMa, and $\theta$ Leo -- have been found to host ultra-weak magnetic fields with strengths $\lesssim1\,{\rm G}$ ($\sim100$ times weaker than the weakest fields found on mCP stars) \citep{lignieres2009,petit2011a,blazere2016}. Low-contrast spots presumably caused by inhomogeneous chemical or temperature distributions have also recently been inferred from spectroscopic observations of Vega \citep{bohm2015,petit2017}. Two important questions related to Vega's spectroscopically detected spots currently remain unanswered: (1) are they directly linked to the star's magnetic structure (i.e. are the spot sizes and locations correlated with Vega's complex magnetic field topology) and (2) do the spots modulate Vega's brightness with its rotation? The discovery of ultra-weak fields on A-type stars in conjunction with the reported discovery of a large sample of A stars that seem to exhibit rotationally modulated light curves has led to speculation that many, or perhaps all, MS A-type stars host detectable magnetic fields (those with strengths $\gtrsim1\,{\rm G}$) \citep{blazere2016,petit2017}.

Assuming that the variability discovered by \citet{balona2013} is intrinsic to the A stars themselves, it is plausible that it has a non-magnetic origin (i.e. it is not produced by or linked to the presence of either strong or ultra-weak magnetic fields). For example, within the last two decades, a small number of HgMn stars have been found to exhibit (i) spectral line variability \citep{adelman2002,kochukhov2005,folsom2010} and (ii) photometric variability \citep{alecian2009,balona2011,hummerich2018}, which may be indicative of surface chemical and/or temperature spots. Doppler imaging of HgMn stars has shown that these surface structures are much larger in scale compared to those found on Vega and have been found to evolve with time \citep[e.g.][]{kochukhov2007}. However, essentially all of the high-precision spectropolarimetric surveys that have been carried out in search of weak fields that may be producing such spots in these stars have yielded null results; for example, \citet{auriere2010} and \citet{makaganiuk2011,makaganiuk2012} reported no field detections while achieving longitudinal field uncertainties as low as $\sim1\,{\rm G}$.

In addition to the surprisingly high incidence rate of A stars inferred to exhibit rotational modulation, \citet{balona2013} also reported the detection of flares that appear to be associated with $\approx2$~per~cent of the A stars in the \emph{Kepler} sample. \citet{pedersen2017} carried out a study of 33 of these flaring A stars and concluded that, in at least 19 cases, the detected flares are not intrinsic to the A stars, but rather they are more likely attributable to low-mass companions or contamination from background stars. Whether the detected rotational modulation is intrinsic to the A stars themselves and, if so, whether it is produced by chemical and/or temperature spots are important unanswered questions, particularly in the context of stellar magnetism. \citet{balona2017} and more recently, \citet{balona2019b}, have begun to address the first of these questions using published projected rotational velocities ($v\sin{i}$). These two studies identified 223 MS A- and late B-type stars with effective temperatures of $8\,000-15\,000\,{\rm K}$ that (1) appear to exhibit rotationally modulated \textit{Kepler}, \textit{K2}, or \textit{TESS} light curves and (2) have available $v\sin{i}$ values. It was concluded that the majority of this subset have $v\sin{i}\lesssim v_{\rm eq}$ where $v_{\rm eq}$ is each star's equatorial velocity derived under the assumption that the identified photometric period corresponds to the rotation period \citep[e.g. Fig. 1 of][]{balona2019b}.

In the following, we present the results of a spectroscopic survey of a sample of MS A-type stars identified by \citet{balona2013} as exhibiting rotationally modulated \emph{Kepler} light curves. The goals of this survey are (i) to test the rotational modulation hypothesis by measuring each star's $v\sin{i}$ value and comparing with $v_{\rm eq}$ inferred under the assumption that the photometric periods are the rotation periods, (ii) to search for line profile variability that may be a diagnostic of surface structures similar to that associated with Ap/Bp stars, (iii) to search for strong chemical peculiarities that may provide insight into the origin of the observed variability, and (iv) to search for the presence of low-mass binary companions, where such detections may provide an alternative explanation for the variability (i.e. the variability may be intrinsic to such companions or it may be associated with orbital motions or tidal distortions).

In Sections \ref{sect:sample} and \ref{sect:obs} we introduce the sample of stars included in our survey along with the observations used in this study. The derivation of each star's fundamental parameters, chemical abundances, and radial velocities derived from photometric and spectroscopic measurements are presented in Sect. \ref{sect:fund_param}. In Sect. \ref{sect:binarity}, we present our search for radial velocity variability that is indicative of binary companions. Sect. \ref{sect:phot_var} presents a new analysis of the \emph{Kepler} light curves that exhibit variability that is hypothesized to be due to rotational modulation. In Sect. \ref{sect:test}, we compare the derived rotational broadening parameters with implied equatorial velocities in order to test whether the periods inferred from the \emph{Kepler} light curves can plausibly be attributed to each star's rotation period. Finally, in Sections \ref{sect:discussion} and \ref{sect:conclusion} we discuss the results of our survey and the conclusions that can be drawn.

\section{Sample selection}\label{sect:sample}

Our sample was selected from the list of 875 targets identified by \citet{balona2013} as exhibiting rotational modulation in their \emph{Kepler} light curves. The full sample of stars that \citet{balona2013} searched for such signatures consisted of all \emph{Kepler} targets with effective temperatures ($T_{\rm eff}$) listed in the \emph{Kepler} Input Catalogue (KIC) \citep{brown2011} as being between $7\,500$ and $10\,000\,{\rm K}$ ($1\,974$ stars). This $T_{\rm eff}$ range was adopted in order to approximately select all stars with A9 to A0 spectral types. The $T_{\rm eff}$ values reported in the KIC are derived from photometric observations obtained primarily using SDSS $griz$ filters. \citet{brown2011} note that they were unable to obtain $u$ observations for all of the stars within the \emph{Kepler} field due to the prohibitively long exposure times that were required and, as a result, they consider the effective temperatures included in the KIC with values $\geq9\,000\,{\rm K}$ to be unreliable.

Based on the list of 875 rotationally variable A-type stars identified by \citet{balona2013}, we selected those targets with $V<9\,{\rm mag}$. This magnitude limit was adopted because it yielded a reasonable sample size of 44 stars for which multiple moderate-signal-to-noise (S/N) spectroscopic observations could be obtained per star while also maintaining relatively low exposure times. The exposure times were calibrated in such a way as to achieve a target ${\rm S/N}\sim100$~per~pixel at a wavelength of $\approx5\,500\,{\rm \AA}$, which we estimated to be adequate for the derivation of radial velocities to a precision of $<5\,{\rm km\,s}^{-1}$. Based on the spectral types that were found in the literature, 23 of the 44 stars in our sample are late-B (B7-B9) or early-A (A0-A2) type stars; the majority of the 27 stars in the sample with published luminosity classes are identified as MS stars (class IV or V) while 3 are identified as being evolved (class III). One of the 44 sample stars is a mid-B type star, 9 are mid- to late-A (A3-A9) type stars, and no spectral types could be found for the remaining 11 stars. The sample also contains two known Am stars (KIC~8692626 and KIC~8703413), one known Bp star (KIC~8324268), and two known Be stars (KIC~7131828 and KIC~3848385). The extracted spectral types are listed in Table \ref{tbl:obs}.

\begin{table*}
	\caption[The sample of 44 A- and B-type stars included in this study.]{The sample of 44 A- and B-type stars included in this study. Columns 1 to 4 list each target's KIC number, HD number, spectral type, and $V$ magnitude. Column 5 lists the number of Stokes $I$ observations obtained with ESPaDOnS for each star. Columns 6 and 7 list the maximum and median achieved S/N per pixel at a wavelength of $5\,500\,{\rm \AA}$ associated with the observations while column 8 lists the S/N per pixel associated with each star's averaged spectrum. Columns 9 and 10 list the minimum and maximum time between any two observations.}
	\label{tbl:obs}
	\begin{center}
	\begin{tabular}{@{\extracolsep{\fill}}l c c c c c c c c r@{\extracolsep{\fill}}}
		\hline
		\hline
		\noalign{\vskip0.5mm}
				KIC  & HD   & Sp.  & $V$   & No.  & Max                   & Median                & Combined              & $\Delta t_{\rm min}$ & $\Delta t_{\rm max}$ \\
				ID   & num. & Type & (mag) & obs. & S/N$\,{\rm pxl}^{-1}$ & S/N$\,{\rm pxl}^{-1}$ & S/N$\,{\rm pxl}^{-1}$ & (d)                  & (d)                  \\
				(1)  & (2)  & (3)  & (4)   & (5)  & (6)                   & (7)                   & (8)                   & (9)                  & (10)                 \\
		\noalign{\vskip0.5mm}
		\hline
		\noalign{\vskip0.5mm}
1572201      &  182757 &                               &    8.54 &  7 &  116 &   95 &  254 &       1.0 &       285 \\
2859567      &  184217 &          B9.5IV-V$^{\rm \,a}$ &    8.21 &  7 &  121 &   95 &  249 &       1.0 &       285 \\
3629496      &  177877 &             B8.5V$^{\rm \,a}$ &    8.21 &  8 &  119 &  106 &  293 &       0.9 &       264 \\
3848385      &  182550 &               B8V$^{\rm \,b}$ &    8.94 &  7 &  119 &  100 &  262 &       1.0 &       289 \\
4048716      &  180914 &               A1V$^{\rm \,c}$ &    8.43 &  8 &  117 &  103 &  282 &       0.9 &       291 \\
4567097      &  184469 &                B9$^{\rm \,d}$ &    7.75 &  7 &  126 &   96 &  271 &       1.0 &       285 \\
4663468      &  185240 &                               &    8.68 &  7 &  123 &  109 &  257 &       1.0 &       281 \\
4818496      &  177592 &               A1V$^{\rm \,c}$ &    8.07 &  8 &  120 &   98 &  264 &       1.0 &       265 \\
4829781      &  181779 &              B9IV$^{\rm \,e}$ &    8.93 &  7 &  118 &   98 &  258 &       1.0 &       289 \\
4995049      &  177982 &                               &    8.53 &  8 &  111 &  108 &  277 &       1.0 &       264 \\
5371784      &  185525 &                               &    8.53 &  6 &  117 &  106 &  255 &       1.0 &       144 \\
5395418      &  226874 &                A0$^{\rm \,c}$ &    8.70 &  7 &  110 &   90 &  232 &       1.0 &       175 \\
5430514      &  177081 &                               &    8.89 &  7 &  114 &  105 &  268 &       1.0 &       264 \\
5436432      &  179618 &                A2$^{\rm \,f}$ &    8.95 &  7 &  115 &  105 &  259 &       1.0 &       264 \\
5461344      &  186254 &             B6III$^{\rm \,e}$ &    8.60 &  7 &  117 &  106 &  263 &       1.0 &       175 \\
5880360      &  184380 &            A3IVan$^{\rm \,g}$ &    8.75 &  7 &  110 &   92 &  235 &       1.0 &       281 \\
6106152      &  177061 &       A3IVwkmetA1$^{\rm \,g}$ &    8.06 &  9 &  115 &  106 &  289 &       0.9 &       265 \\
6450107      &  185265 &             A1IVs$^{\rm \,h}$ &    7.53 &  7 &  125 &  107 &  257 &       1.0 &       281 \\
7050270      &  186883 &                               &    8.74 &  7 &  120 &  106 &  278 &       1.0 &       175 \\
7131828      &  186485 &             B9Vne$^{\rm \,e}$ &    8.52 &  6 &  119 &  103 &  252 &       1.0 &       175 \\
7345479      &  177328 &             A2Vnn$^{\rm \,g}$ &    7.90 &  8 &  120 &  109 &  287 &       0.9 &       265 \\
7383872      &  187710 &                               &    8.50 &  7 &  110 &   97 &  261 &       0.9 &       179 \\
7530366      &  184024 &          A0.5IVnn$^{\rm \,h}$ &    8.31 &  7 &  122 &  100 &  263 &       1.0 &       285 \\
7974841      &  187139 &             A0III$^{\rm \,i}$ &    8.16 &  7 &  117 &  105 &  261 &       0.9 &       179 \\
8153795      &  178847 &               B9V$^{\rm \,j}$ &    7.87 &  9 &  130 &  105 &  302 &       0.9 &       265 \\
8324268      &  189160 &             B8pSi$^{\rm \,c}$ &    7.90 &  7 &  118 &  101 &  263 &       0.9 &       179 \\
8351193      &  177152 &         A0VkB8mB7$^{\rm \,i}$ &    7.57 &  9 &  124 &  106 &  304 &       0.9 &       265 \\
8367661      &  184023 &             A2.5V$^{\rm \,i}$ &    8.56 &  7 &  119 &   97 &  253 &       1.0 &       281 \\
8390826      &  189375 &                               &    8.97 &  6 &   99 &   92 &  206 &       1.0 &       175 \\
8692626      &  184482 &       kA2hA4mA6IV$^{\rm \,h}$ &    8.29 &  7 &  120 &   93 &  259 &       1.0 &       285 \\
8703413      &  187254 &        kA5hA5mF2V$^{\rm \,i}$ &    8.71 &  6 &  112 &  100 &  244 &       1.0 &       175 \\
9349245      &  185658 &             A7III$^{\rm \,e}$ &    8.11 &  6 &  113 &  104 &  252 &       1.0 &       144 \\
9392839      &  177931 &                B9$^{\rm \,k}$ &    7.18 &  8 &  123 &  112 &  301 &       0.9 &       265 \\
9468475      &  184602 &               A4V$^{\rm \,e}$ &    7.58 &  7 &  115 &  102 &  254 &       0.9 &       281 \\
9772586      &  184086 &               A2V$^{\rm \,e}$ &    8.96 &  7 &  115 &   93 &  246 &       1.0 &       281 \\
10724634     & 181094A &                               &    8.82 &  7 &  116 &  104 &  262 &       1.0 &       291 \\
10815604     &  188360 &                A0$^{\rm \,d}$ &    8.06 &  7 &  119 &  103 &  271 &       0.9 &       179 \\
10879812     &  188610 &                               &    8.12 &  6 &  111 &  101 &  241 &       1.0 &       175 \\
10974032     &  182828 &               A0V$^{\rm \,i}$ &    8.30 &  9 &  116 &  101 &  290 &       0.9 &       289 \\
11189959     &  183257 &              A1Vb$^{\rm \,i}$ &    8.15 &  7 &  125 &  112 &  267 &       1.0 &       285 \\
11443271     &  176708 &               A2V$^{\rm \,i}$ &    7.46 &  8 &  135 &  118 &  296 &       1.0 &       265 \\
11600717     &  177828 &               A8V$^{\rm \,i}$ &    7.53 &  9 &  117 &  103 &  298 &       0.9 &       265 \\
12061741     &  183254 &               A1V$^{\rm \,i}$ &    8.45 &  7 &  123 &  107 &  270 &       1.0 &       285 \\
12306265     &  182613 &                               &    8.59 &  7 &  117 &  105 &  269 &       1.0 &       289 \\
		\noalign{\vskip0.5mm}
		\hline \\
\noalign{\vskip-0.3cm}
\multicolumn{10}{l}{$^{\rm a\,}$\citet{tkachenko2013}, $^{\rm b\,}$\citet{sato1990}, $^{\rm c\,}$\citet{skiff2014}, $^{\rm d\,}$\citet{davis1973}} \\
\multicolumn{10}{l}{$^{\rm e\,}$\citet{frasca2016}, $^{\rm f\,}$\citet{kharchenko2001a}, $^{\rm g\,}$\citet{niemczura2017}, $^{\rm h\,}$\citet{niemczura2015}} \\
\multicolumn{10}{l}{$^{\rm i\,}$\citet{pedersen2017}, $^{\rm j\,}$\citet{grillo1992}, $^{\rm k\,}$\citet{wilson1953}} \\
	\end{tabular}
	\end{center}
\end{table*}

\section{Observations}\label{sect:obs}

This study primarily makes use of two observational data sets: \emph{Kepler} photometric light curves and high-resolution spectroscopic measurements obtained with ESPaDOnS at Canada France Hawaii Telescope (CFHT).

\subsection{\emph{Kepler} photometric measurements}\label{sect:obs_phot}

The photometric variability associated with the 44 stars in our sample was first reported by \citet{balona2013} based on light curves obtained with the \emph{Kepler} spacecraft \citep{borucki2010}. The passband of the filter used for these photometric measurements has an effective wavelength of $5\,800\,{\rm \AA}$ and an effective width of $\approx3\,500\,{\rm \AA}$. The angular size of \emph{Kepler}'s CCD pixels is $4"$ \citep{koch2010}.

All of the \emph{Kepler} observations used for the 44 stars in this study were obtained from the Mikulski Archive for Space Telescopes (MAST)\footnote{https://archive.stsci.edu/kepler/} following data release 25. We used all of the available long cadence light curves (i.e. $\Delta t=30\,{\rm min}$), which were taken over a time period of $4\,{\rm yrs}$ from May 2, 2009 to May 11, 2013. The majority of the available light curves (41 of the 44 stars) span time periods between 13 and 18 observing quarters, where each quarter spans approximately $3\,$~months; two of the 44 stars have light curves spanning 6 and 9 quarters (KIC~3629496 and KIC~3848385, respectively) while measurements spanning only a single quarter (quarter 3) are available for KIC~11600717.

We used the PDC\_SAP light curves, which have been processed by the \emph{Kepler} team to correct for various errors such as the removal of outliers and systematic trends \citep{smith2012}. All PDC\_SAP flux measurements that were stored as `NaN' were removed. We carried out additional post-processing, which involved removing (i)  remaining outliers and (ii) long-term trends occuring over time scales on the order of each observing quarter; the de-trending was carried out by fitting first- or second-order polynomials to the flux measurements spanning each quarter and dividing by the resulting fit.

\subsection{ESPaDOnS spectroscopic measurements}\label{sect:obs_spec}

The Stokes $I$ spectroscopic measurements presented in this study were obtained using the ESPaDOnS \'echelle spectrograph installed at the $3.6\,{\rm m}$ CFHT. This instrument has a high spectral resolving power ($R\sim65\,000$) and is optimized for a wavelength range of approximately $3\,600-10\,000\,{\rm \AA}$. All of the measurements were reduced using the Upena pipeline feeding the {\sc libre-esprit} reduction code described by \citet{donati1997}.

One of the goals of this study is to detect radial velocity variability induced by the presence of binary companions. Given that the orbital periods of any such binary companions are unknown, we needed to obtain multiple measurements of each sample star's radial velocity over both short ($\lesssim1\,{\rm d}$) and long ($\gtrsim100\,{\rm d}$) timescales. We adopted the observing strategy employed by \citet{lampens2018} in which the time between consecutive observations is gradually increased from $\lesssim1\,{\rm d}$, to $\sim7\,{\rm d}$, and finally to $\sim30\,{\rm d}$. A total of 319 Stokes $I$ observations were obtained over a time period spanning $\approx11\,{\rm months}$ from Jan. 31, 2018 to Dec. 21, 2018; each of the 44 stars in our sample was observed between 6 and 9 times. Eighty-five per cent of the 319 observations have ${\rm S/N}>75$~per~pixel while the median S/N is 100~per~pixel. The observations are summarized in Table \ref{tbl:obs} where we list the total number of times each target was observed, the maximum and median achieved S/Ns, the S/N associated with each star's averaged spectrum, and the minimum and maximum time ($\Delta t_{\rm min}$ and $\Delta t_{\rm max}$, respectively) between any two observations.

\section{Fundamental parameters}\label{sect:fund_param}

We derived fundamental parameters for the stars in our sample (effective temperatures, surface gravities, metallicities, radii, masses, etc.). This was carried out using both the high-resolution spectroscopic observations obtained for this study and published multi-colour photometric observations obtained using various filters.

\subsection{Spectral fitting}\label{sect:spec_fit}

Spectral modelling of the Stokes $I$ ESPaDOnS observations was carried out using the Grid Search in Stellar Parameters ({\sc gssp}) code \citep{tkachenko2015}. This code uses grids of pre-computed {\sc LLmodels} atmospheric models \citep{shulyak2004} in order to generate grids of synthetic spectra that span a range of parameters. The parameters that can be varied include effective temperature ($T_{\rm eff}$), surface gravity ($\log{g}$), metallicity ([M/H]), rotational broadening ($v\sin{i}$), and microturbulence ($\xi$). Individual chemical abundances can also be fit one element at a time. The $\chi^2$ values associated with each model in the defined grid are computed and the minimal $\chi^2$ model can be determined along with each parameter's uncertainty based on the associated $\chi^2$ intervals.

\begin{figure*}
	\centering
	\subfigure{\includegraphics[width=1.5\columnwidth]{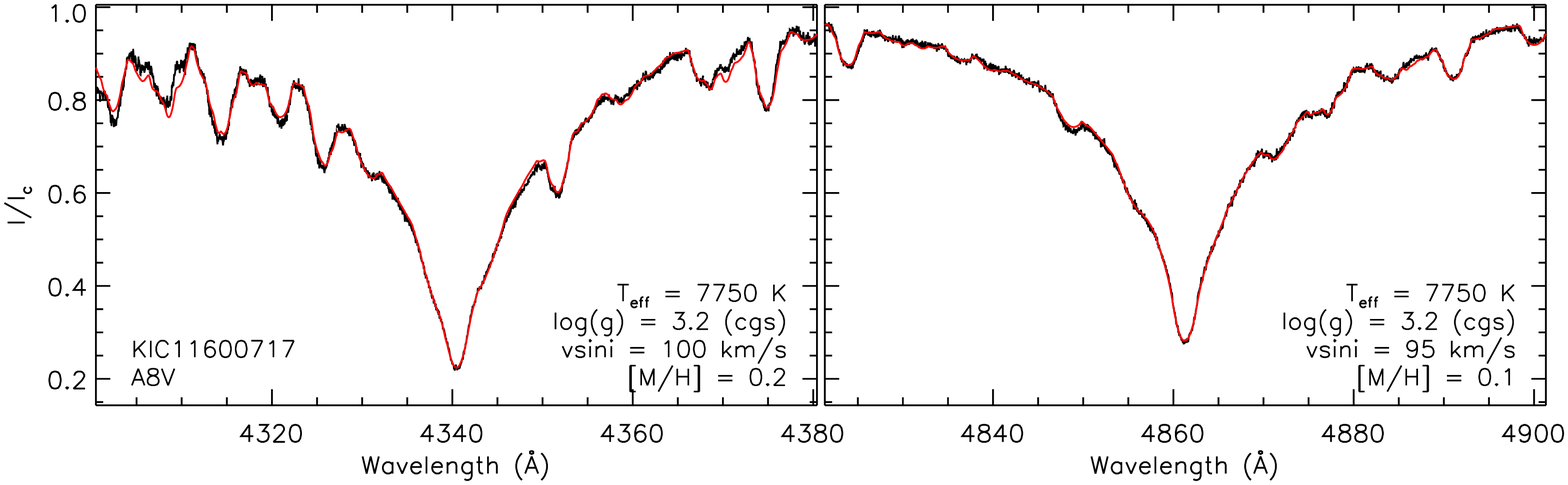}}\vspace{-0.4cm}
	\subfigure{\includegraphics[width=1.5\columnwidth]{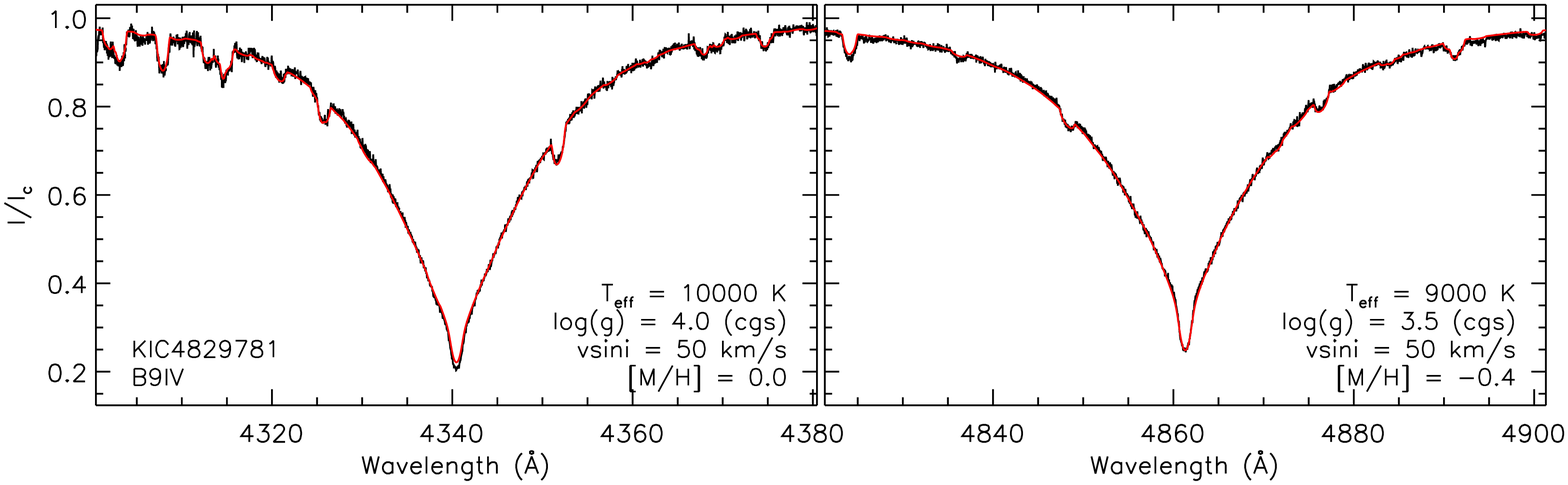}}
	\caption[Examples of the fits to the observed average H$\gamma$ and H$\beta$ profiles.]{Examples of the fits to the observed average H$\gamma$ and H$\beta$ profiles. The black curves correspond to the observations while the red curve corresponds to the synthetic spectra.}
	\label{fig:Balmer_fit}
\end{figure*}

The spectral modelling analysis consisted of the following three steps.
\begin{enumerate}
	\item We derived $T_{\rm eff}$, $\log{g}$, [M/H], and $v\sin{i}$ values by fitting H$\beta$ and H$\gamma$ line profiles, which are both relatively sensitive to changes in $T_{\rm eff}$ and $\log{g}$ for A and late-B type stars. KIC~3848385, KIC~5371784, and KIC~7131828 are Be stars and exhibit significant emission within their H$\beta$ and H$\gamma$ line profiles (these cases are discussed in more detail in Sect. \ref{sect:balmer_emission}) and as a result, their Balmer lines could not be modelled using {\sc gssp}. In these three cases, we used narrow $21\,{\rm \AA}$-width spectral windows centered at $4\,475\,{\rm \AA}$ to derive $T_{\rm eff}$, $\log{g}$, [M/H], and $v\sin{i}$ values. This region was selected because of the fact that it encompasses various spectral lines that are, to various degrees, sensitive to both $T_{\rm eff}$ and $\log{g}$: for A-type stars with $7\,000\lesssim T_{\rm eff}\lesssim10\,000\,{\rm K}$ this spectral region includes Fe~{\sc i}, Fe~{\sc ii}, and Mg~{\sc ii} lines while for early-A and late-B type stars with $10\,000\lesssim T_{\rm eff}\lesssim15\,000\,{\rm K}$ this region includes He~{\sc i} and Mg~{\sc ii} lines. We note that the $21\,{\rm \AA}$-width spectral window was found to be less sensitive to changes in $T_{\rm eff}$ and $\log{g}$ compared to H$\beta$ and H$\gamma$, which is demonstrated by the typically larger uncertainties yielded by the {\sc gssp} code: the median $T_{\rm eff}$ and $\log{g}$ uncertainties for those three stars exhibiting Balmer line emission are $1\,250\,{\rm K}$ and $1.2\,{\rm (cgs)}$, respectively, while the values derived for the stars not exhibiting Balmer line emission are $400\,{\rm K}$ and $0.4\,{\rm (cgs)}$, respectively. The microturbulence was fixed at $\xi=2\,{\rm km\,s}^{-1}$ during this first step of the analysis.
	\item Multiple $100\,{\rm \AA}$-width spectral windows ranging from $4\,900\,{\rm \AA}$ to $5\,900\,{\rm \AA}$ were then fit in order to place additional constraints on [M/H] and $v\sin{i}$ and to derive $\xi$; $T_{\rm eff}$ and $\log{g}$ were fixed at the values derived during the first step of the analysis.
	\item Finally, the $100\,{\rm \AA}$-width spectral windows used in the previous step were fit while varying individual chemical abundances with all other parameters fixed.
\end{enumerate}

Prior to fitting, the observed spectra were normalized using two methods depending on whether or not Balmer lines were being fit. In both cases, we used the un-normalized spectra that were reduced with {\sc libre-esprit} \citep{donati1997}. The normalization of the H$\beta$ and H$\gamma$ profiles was carried out by selecting narrow wavelength regions spanning $\sim1\,{\rm \AA}$ on either side of the Balmer line's central wavelength ($\lambda_0$). These regions have wavelengths centered at approximately $\lambda_0\pm40\,{\rm \AA}$ and are therefore outside of the Lorentzian-broadened region of the Balmer line profiles. A first order polynomial function was then fit across the line within the selected regions and the entire Balmer line profile was normalized to the resulting fit. The spectral windows that do not include Balmer lines were normalized using a polynomial fit -- typically of $1^{\rm st}$ or $2^{\rm nd}$ degree -- to the continuum.

Initial parameter grids used during the execution of {\sc gssp} were centered on $T_{\rm eff}=T_{\rm eff}^{\rm KIC}$ \citep[where $T_{\rm eff}^{\rm KIC}$ is the effective temperature listed in the Kepler Input Catalog;][]{brown2011}, $\log{g}=4.0\,{\rm (cgs)}$, ${\rm [M/H]}=0.0$, and $\xi=2.0\,{\rm km\,s}^{-1}$ while initial $v\sin{i}$ values were estimated by eye. These initial grids spanned $1\,000\,{\rm K}$ in $T_{\rm eff}$, $0.6\,{\rm (cgs)}$ in $\log{g}$, $0.5$ in [M/H], $4.0\,{\rm km\,s}^{-1}$ in $\xi$, and $50\,{\rm km\,s}^{-1}$ in $v\sin{i}$. We adopted grid resolutions of $\Delta T_{\rm eff}=100\,{\rm K}$, $\Delta\log{g}=0.1\,{\rm (cgs)}$, $\Delta({\rm [M/H]})=0.1$, $\Delta\xi=0.5\,{\rm km\,s}^{-1}$, and $\Delta v\sin{i}=1\,{\rm km\,s}^{-1}$. After each execution of the {\sc gssp} code, the parameter grids were re-centered on the best-fitting values and, if necessary, the range of the grids was expanded until either (i) upper and lower $1\sigma$ confidence limits were reached or (ii) either physically plausible limits were reached (e.g. we expect MS A- and late B-type stars to have $5\,000<T_{\rm eff}<20\,000$, $\log{g}>3.0\,{\rm [cgs]}$, etc.) or the limits of the grid of {\sc LLmodels} atmospheric models were reached (i.e. $T_{\rm eff}\in[5\,000,20\,000]\,{\rm K}$, $\log{g}\in[3.0,5.0]\,{\rm (cgs)}$, [M/H]$\in[-0.8,0.8]$, $\xi\in[0,20]\,{\rm km\,s}^{-1}$, $v\sin{i}\geq0\,{\rm km\,s}^{-1}$). For those executions of the {\sc gssp} codes requiring large parameter ranges, the computation time was slightly reduced for practical purposes by decreasing the grid resolution. In Fig. \ref{fig:Balmer_fit} we show several examples of the obtained fits to the H$\beta$ and H$\gamma$ lines.

The fitting of individual chemical abundances described in step 3 of the spectral modelling analysis was carried out by first identifying those elements that exhibit sufficiently deep lines within the $100\,{\rm \AA}$-width spectral window being fit. This was done by referring to spectral line lists provided by the Vienna Atomic Line Database (VALD) for specified $T_{\rm eff}$ and $\log{g}$ values \citep{piskunov1995,ryabchikova2015}. Any elements found to have lines with unbroadened normalized depths $\geq0.05$ were selected to be fit. Initial abundances were assigned based on the derived [M/H] values. For each element, the {\sc gssp} routine was carried out twice: once with a wide and low-resolution grid ([X/H]$\in[-20,-2]$ and $\Delta\,{\rm [X/H]}=1.0$) and a second time using a narrower, higher-resolution [X/H] grid ($\Delta{\rm [X/H]}=0.1$) centered on the minimum $\chi^2$ value obtained from the first iteration. This was done in order to reduce the total computation time involved in the analysis.

\begin{figure}
	\centering
	\subfigure{\includegraphics[width=0.8\columnwidth]{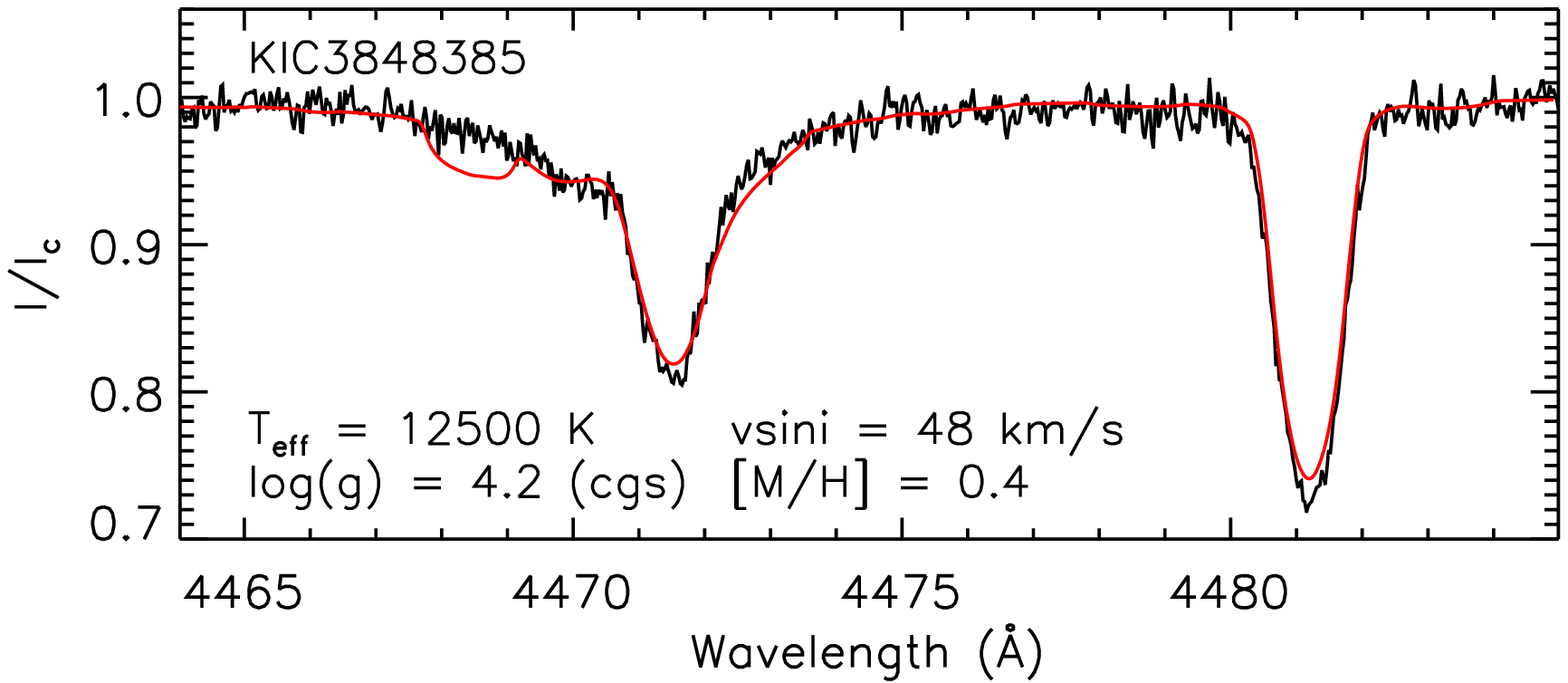}}\vspace{-0.35cm}
	\subfigure{\includegraphics[width=0.8\columnwidth]{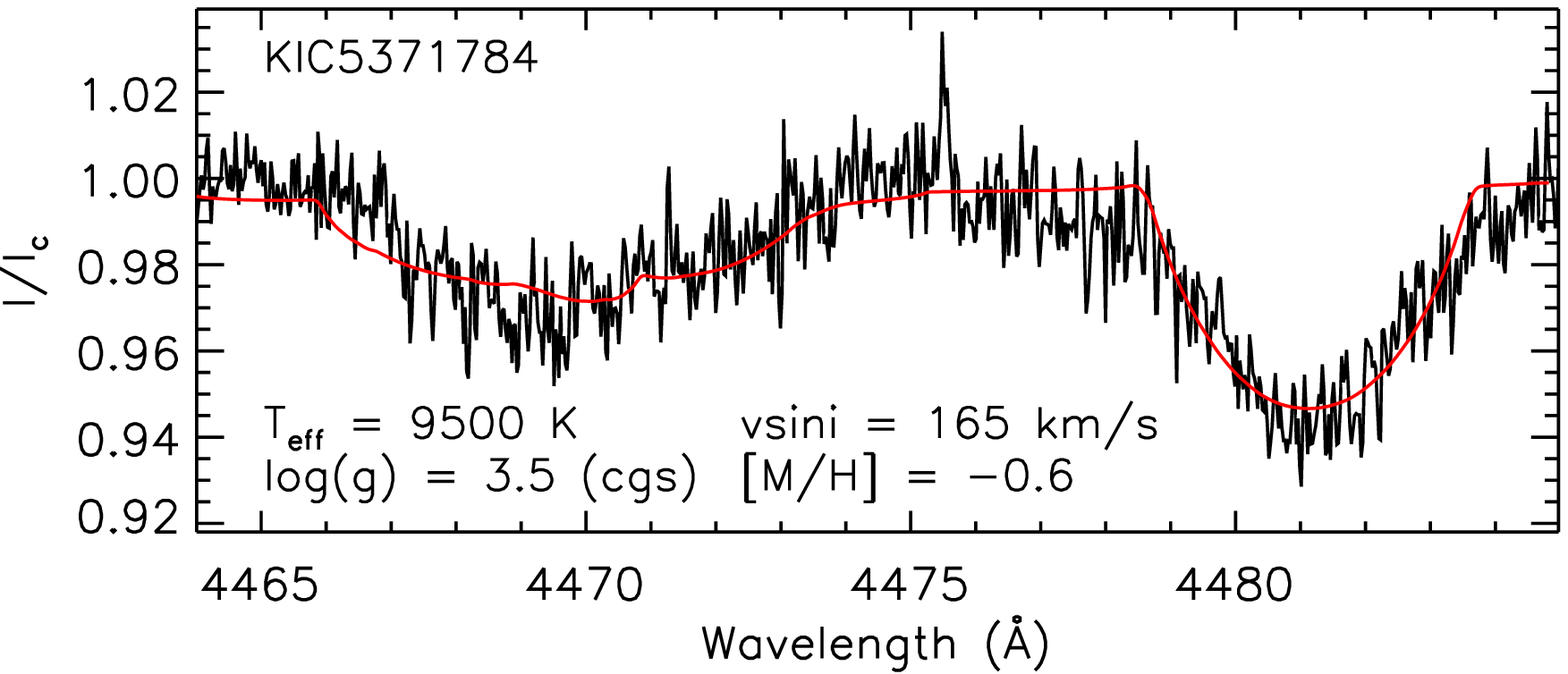}}\vspace{-0.35cm}
	\subfigure{\includegraphics[width=0.8\columnwidth]{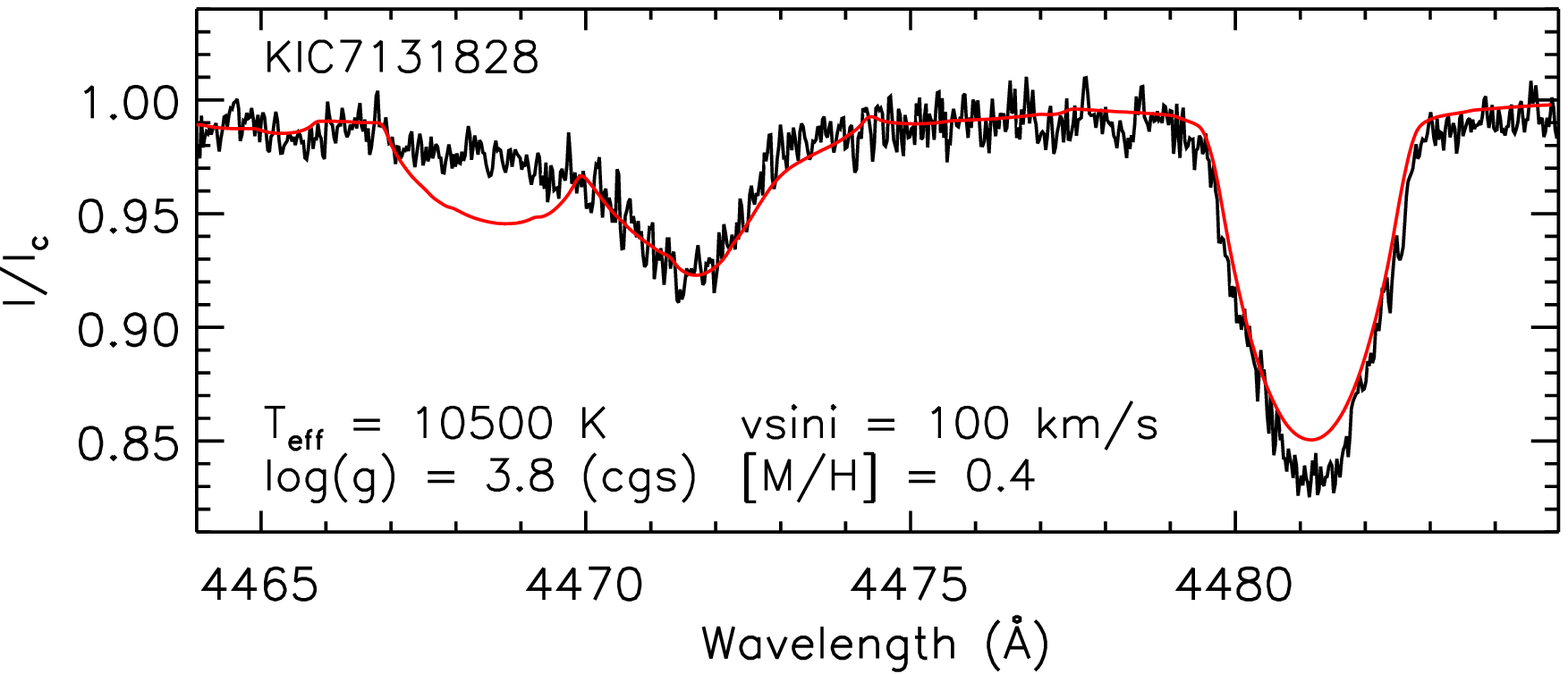}}\vspace{-0.2cm}
	\caption[Fits to the $21\,{\rm \AA}$-width spectral windows centered at $4\,475\,{\rm \AA}$ for the stars in our sample exhibiting strong Balmer line emission.]{Fits to the $21\,{\rm \AA}$-width spectral windows centered at $4\,475\,{\rm \AA}$ for the three stars in our sample that were found to exhibit strong Balmer line emission. As in Fig. \ref{fig:Balmer_fit}, the black curves correspond to the observed average spectra and the red curves correspond to the synthetic spectra. Note that the abundances of individual chemical elements were not varied during this step of the analysis, which likely contributes to the poor fit obtained near $4\,468\,{\rm \AA}$ for KIC~3848385 and KIC~7131828.}
	\label{fig:445_450_fit}
\end{figure}

The final parameters ($T_{\rm eff}$, $\log{g}$, [M/H], $v\sin{i}$, and $\xi$) and chemical abundances derived from the spectral modelling analysis were obtained by fitting each star's averaged observed spectrum. The averaged spectra typically have S/N values two to three times higher than the individual observations (see Table \ref{tbl:obs}). They were obtained by first subtracting any radial velocity shift associated with each individual observed spectrum (the derivation of the radial velocities is discussed below in Sect. \ref{sect:radial_velocities}). The normalized spectra were then interpolated onto a common abscissa and the average flux, weighted by the measurement uncertainties, was computed; care was taken to ensure consistent normalization between the individual spectra. In Fig. \ref{fig:445_450_fit}, we show the fits to the $21\,{\rm \AA}$-width spectral windows centered at $4\,475\,{\rm \AA}$ that were used for the derivation of $T_{\rm eff}$ and $\log{g}$ of the three stars exhibiting Balmer line emission. We compared the best-fitting $T_{\rm eff}$ values derived from the Balmer line profiles ($T_{\rm eff}^{\rm Balmer}$) and from the $4\,475\,{\rm \AA}$ region ($T_{\rm eff}^{\rm 4475}$) for the stars that do not exhibit Balmer line emission. Although the values are nearly all in agreement within the estimated uncertainties, we do see that the $T_{\rm eff}^{\rm 4475}$ values are systematically higher than the $T_{\rm eff}^{\rm Balmer}$ values by $\lesssim500\,{\rm K}$.

\begin{figure}
	\centering
	\subfigure{\includegraphics[width=1\columnwidth]{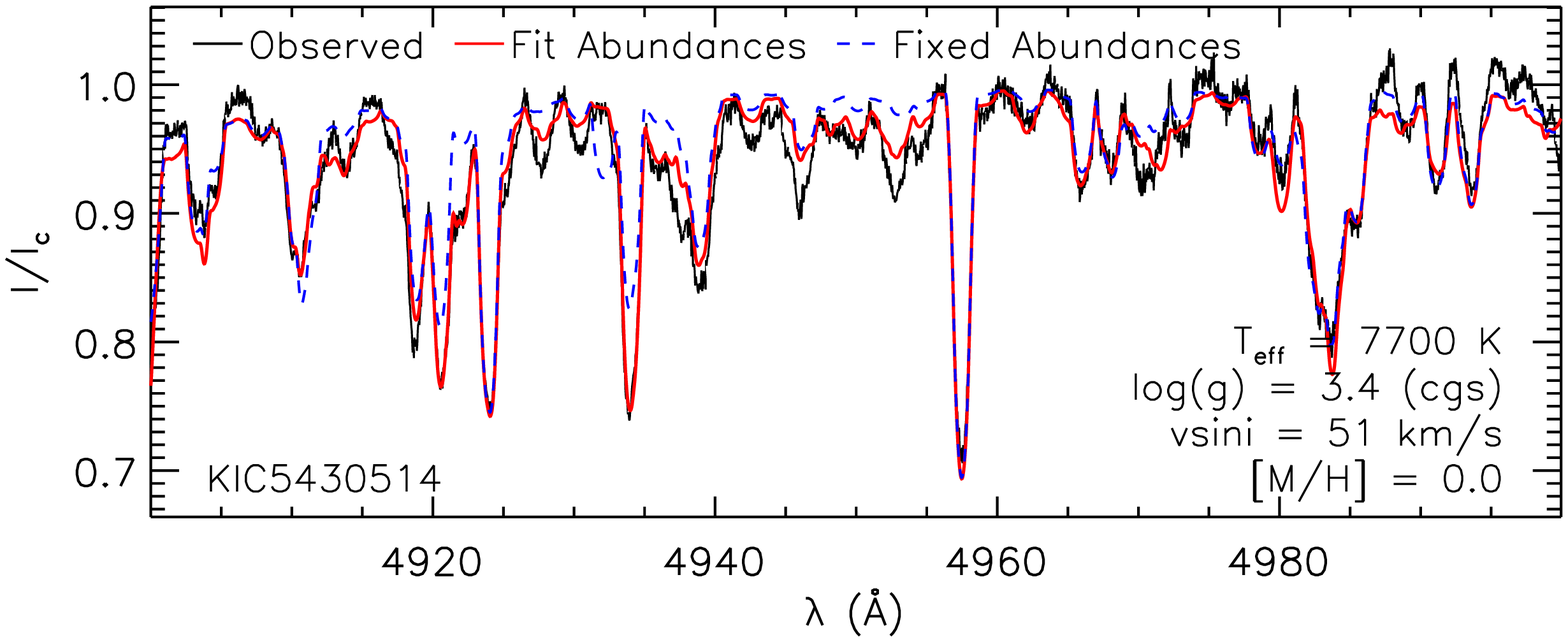}}\vspace{-0.4cm}
	\subfigure{\includegraphics[width=1\columnwidth]{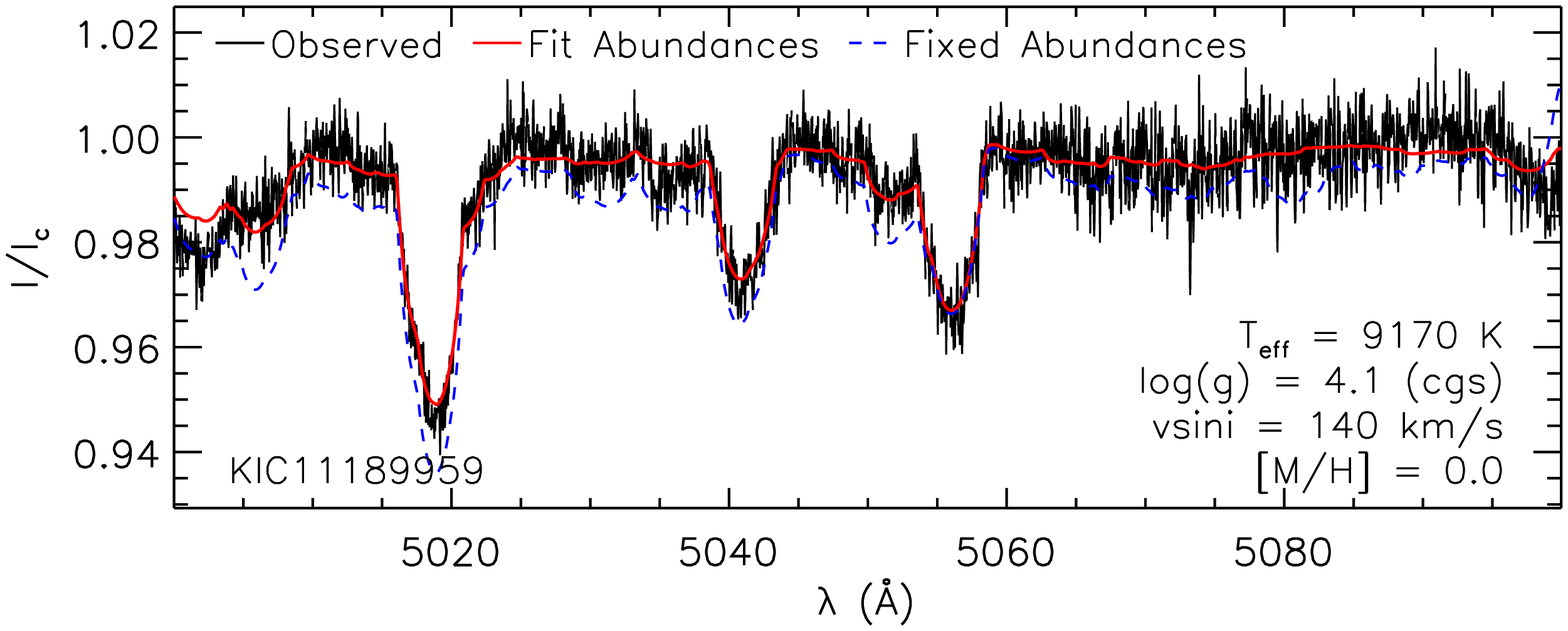}}\vspace{-0.4cm}
	\subfigure{\includegraphics[width=1\columnwidth]{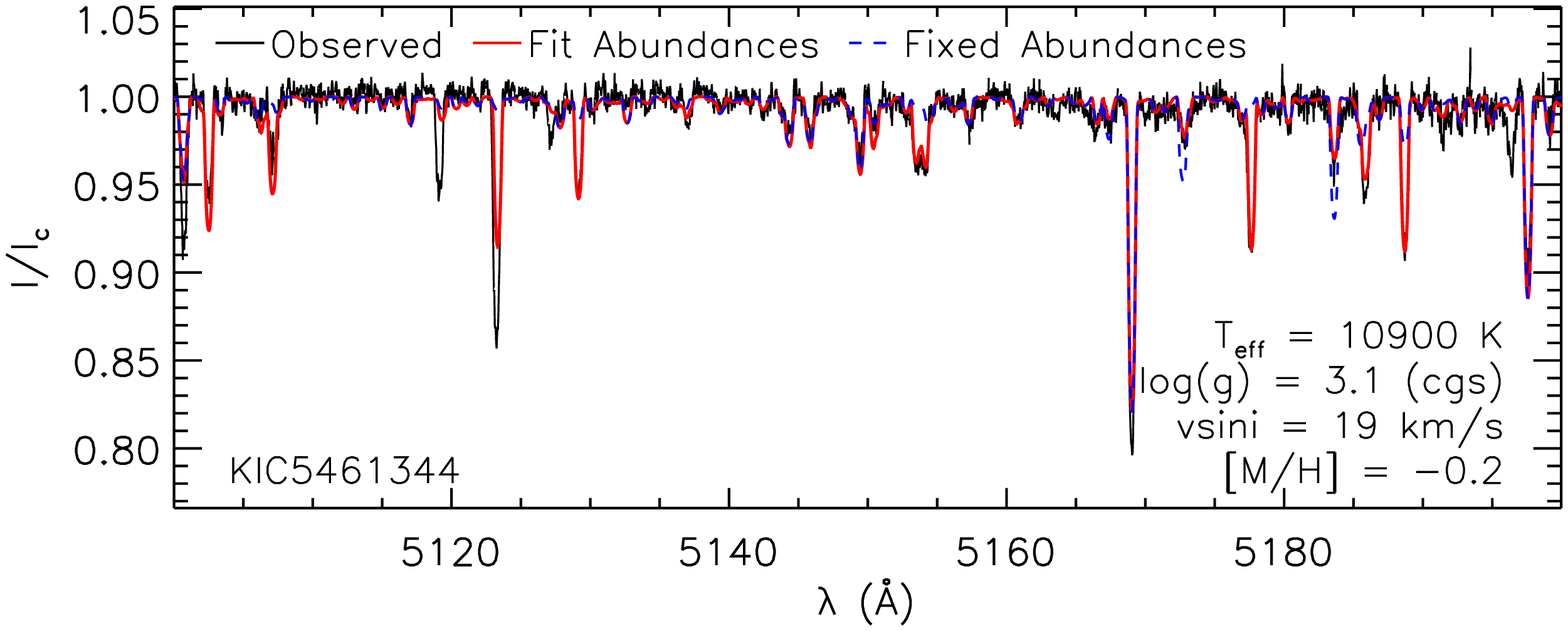}}\vspace{-0.4cm}
	\caption[Examples of fits to the $100\,{\rm \AA}$-width spectral regions derived during step 3 of the spectral modelling analysis.]{Examples of fits to the $100\,{\rm \AA}$-width spectral regions derived during step 3 of the spectral modelling analysis. The black curves correspond to the average observed spectra, the solid-red and dashed-blue curves correspond to the best-fitting synthetic spectra derived with and without fitting individual abundances, respectively.}
	\label{fig:metal_fit}
\end{figure}

The final adopted abundancies were taken to be the weighted average of the values derived by fitting each of the averaged spectral windows where the weights were obtained from the $1\sigma$ $\chi^2$ intervals yielded by the {\sc gssp} code. In the electronic version of this paper, we list the derived $T_{\rm eff}$, $\log{g}$, [M/H], $v\sin{i}$, and $\xi$ values. In three cases, $\xi$ values and individual chemical abundances could not be derived due to low S/Ns and/or high $v\sin{i}$ values, which did not allow for useful constraints to be obtained by fitting the $100\,{\rm \AA}$-width spectral windows. The 41 of 44 stars in our sample for which at least one chemical abundance was derived are discussed below in Sect. \ref{sect:abund}. In Fig. \ref{fig:metal_fit}, we show several examples of the fits to various $100\,{\rm \AA}$-width spectral windows.

\begin{figure*}
	\centering
	\begin{minipage}{0.69\columnwidth}
		\centering
		\subfigure{\includegraphics[width=1.0\columnwidth]{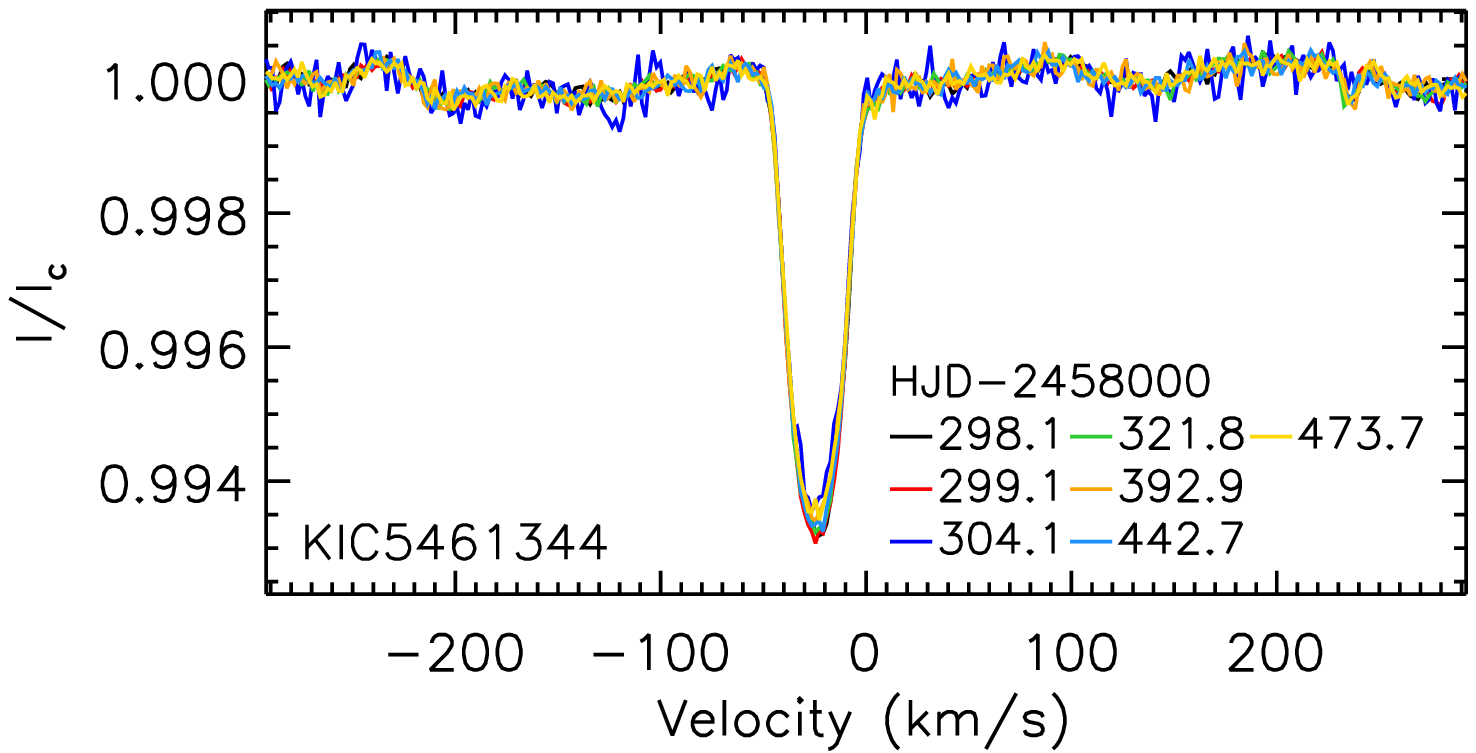}}\vspace{-0.4cm}
		\subfigure{\includegraphics[width=1.0\columnwidth]{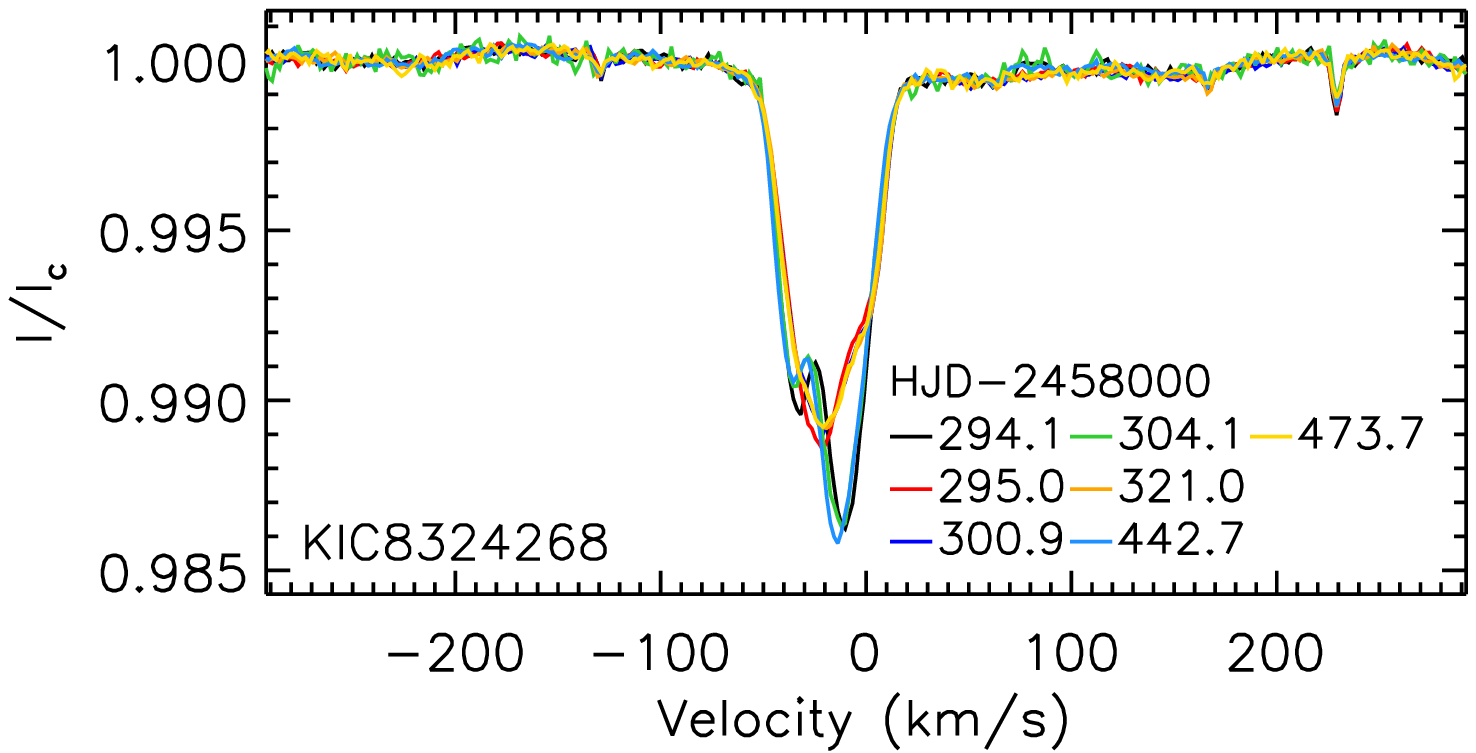}}
	\end{minipage}
	\begin{minipage}{0.69\columnwidth}
		\centering
		\subfigure{\includegraphics[width=1.0\columnwidth]{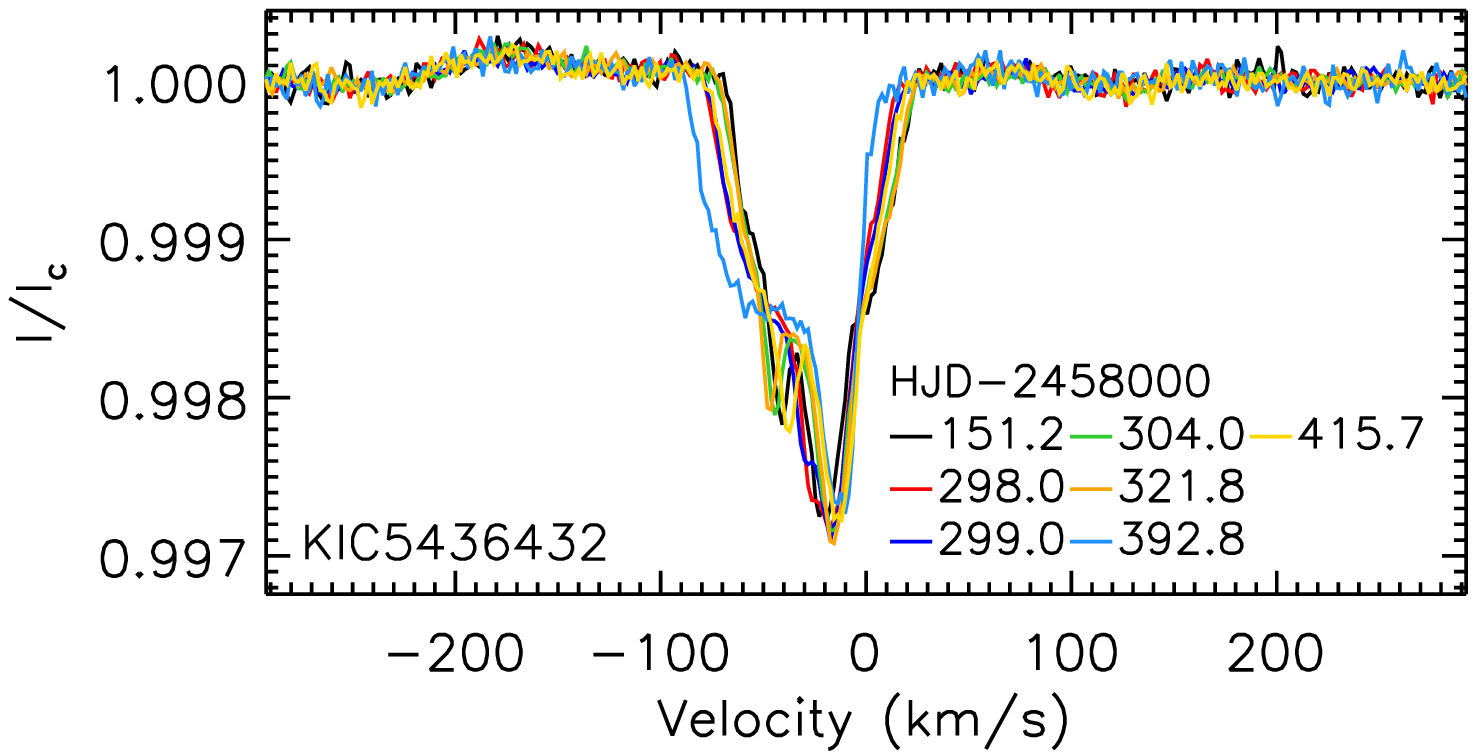}}\vspace{-0.4cm}
		\subfigure{\includegraphics[width=1.0\columnwidth]{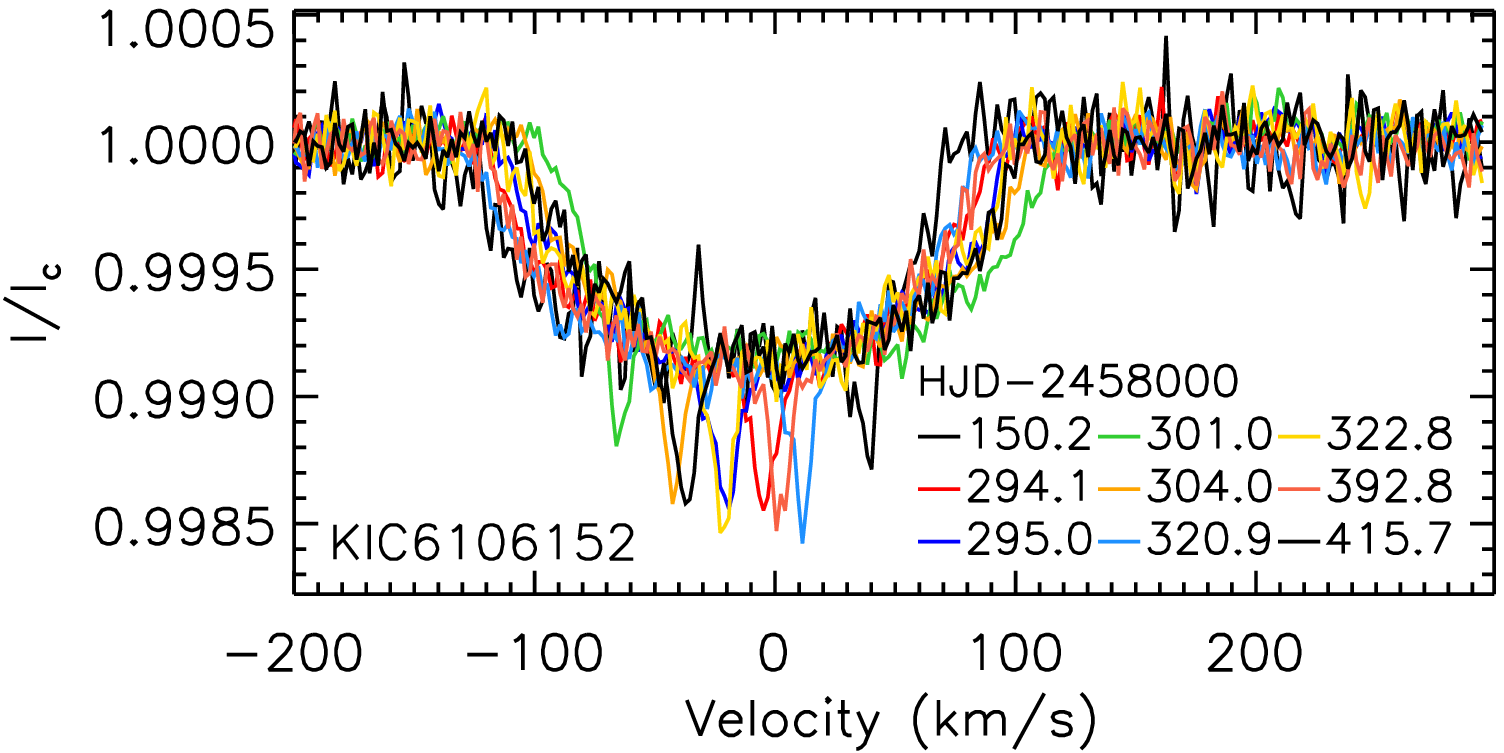}}
	\end{minipage}
	\begin{minipage}{0.69\columnwidth}
		\centering
		\subfigure{\includegraphics[width=1.0\columnwidth]{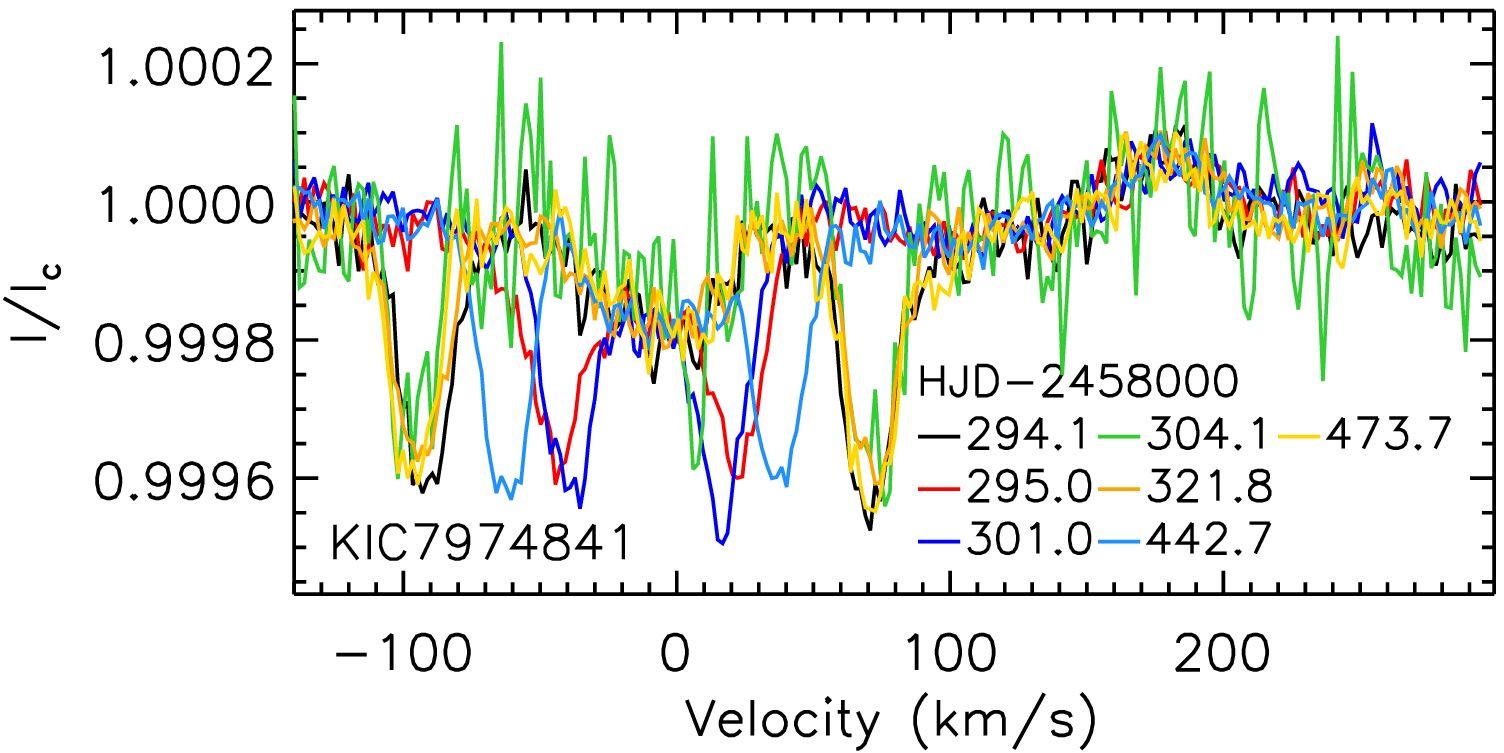}}\vspace{-0.4cm}
		\subfigure{\includegraphics[width=1.0\columnwidth]{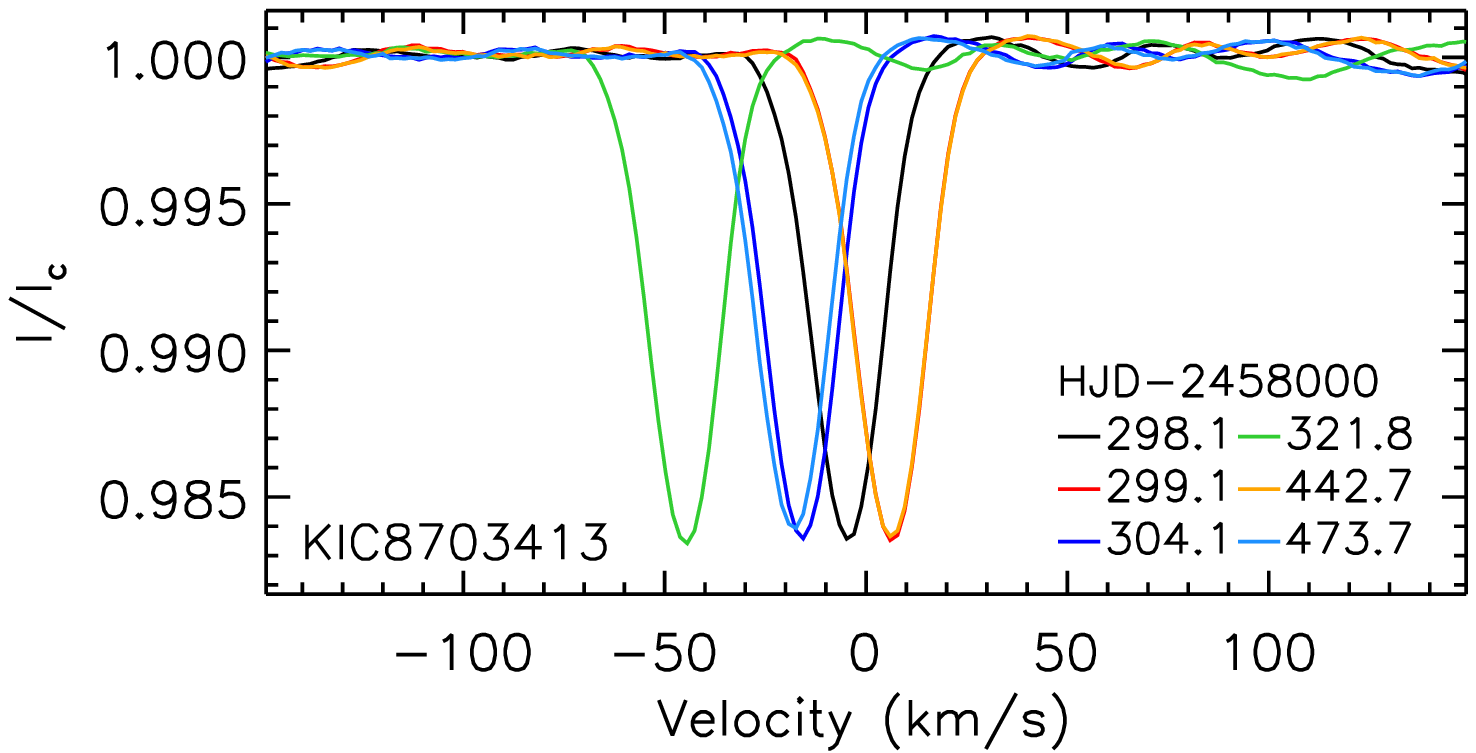}}
	\end{minipage}
	\caption{Examples of those LSD profiles found to exhibit variability that is likely intrinsic to the A stars (left column) or is related to radial velocity variations of one or more stellar components (middle and right columns). The profiles shown for KIC~7974841 were calculated using the $4\,000\,{\rm K}$ line masks, which allow the cooler secondary and tertiary components to be more clearly identified; all of the other LSD profiles that are shown were calculatecd using the A star's derived fundamental parameters.}
	\label{fig:ex_lsd}
\end{figure*}

\subsubsection{Radial velocities}\label{sect:radial_velocities}

We derived the radial velocities using two methods. The first method involved a comparison between the observed spectra and the synthetic models, which was carried out in two steps. First, we calculated synthetic models using the initial parameters ($T_{\rm eff}$, [M/H], $v\sin{i}$, and $\xi$) as described above in Sect. \ref{sect:spec_fit}. These models were then used to roughly estimate the radial velocities by eye. Higher quality spectral fits were subsequently derived using the three-step modelling analysis described in Sect. \ref{sect:spec_fit} while incorporating the roughly estimated radial velocities.

The spectral models that were derived for each of the observed spectra were subsequently used to refine the roughly estimated radial velocities. This involved re-fitting the spectral models to the observed spectra while varying the radial velocity of each model along with two re-normalization parameters. The re-normalization parameters are simply coefficients of a first-order (i.e. linear) polynomial function; the observed spectra are divided by this first-order polynomial in order to minimize any discrepancy that may have been introduced by the initial continuum normalization procedure. We found that including these two re-normalization parameters yielded higher-quality fits and thus, more accurate radial velocities. The fitting analysis was done using the {\sc mpfitfun} Levenberg-Marquardt algorithm implemented in {\sc idl}.

Uncertainties in the radial velocities were estimated by employing a block bootstrapping method \citep[e.g.][]{lahiri2003resampling} involving the best-fitting synthetic model and observed spectra residuals. The models were divided into blocks spanning approximately 5~per~cent of the wavelength range. Equally-sized blocks were then randomly sampled from the residuals and added to the model spectrum blocks; radial velocities were re-derived from these new data sets. This process was repeated 500 times and the uncertainty in each radial velocity measurement was taken to be the standard deviation of the resulting distribution. This method was used rather than simple random sampling with replacement in order to account for correlated errors associated with the models. The final radial velocities and their associated uncertainties were calculated from the weighted average of the individual measurements.

The second method involved generating Least-Squares Deconvolution (LSD) profiles \citep{donati1997,kochukhov2010a}. LSD is a cross-correlation technique in which a large number of spectral lines are combined, resulting in a single line pseudo-profile of higher S/N compared to any individual line. We generated the LSD profiles using line masks consisting of spectral line data provided by VALD \citep{piskunov1995,ryabchikova2015}. Specifically, the line data were obtained using VALD's `Extract Stellar' request in which the star's $T_{\rm eff}$, $\log{g}$, and $\xi$ values are specified along with a wavelength range and a depth threshold. We used the $T_{\rm eff}$ and $\log{g}$ values derived in Sect. \ref{sect:spec_fit}, $\xi=2\,{\rm km\,s^{-1}}$, a wavelength range of $4\,000\,{\rm \AA}$ to $7\,000\,{\rm \AA}$, and a depth threshold equal to 5~per~cent of the continuum. Each line mask was then modified in such a way as to remove all Balmer lines, metallic lines appearing within the broad wings of the Balmer lines, and stellar spectral lines strongly contaminated by telluric lines.

The higher S/Ns obtained using the LSD technique can also be used to detect line profile variability or evidence of additional stellar components that may not be apparent in the spectra (see Fig. \ref{fig:ex_lsd} for examples). For KIC~5461344, KIC~5880360, and KIC~8324268, changes in the line profiles are apparent while no clear radial velocity shifts were detected (based on both a visual inspection and on the variability metric discussed in Sect. \ref{sect:binarity}); therefore, it is plausible that the variability is intrinsic to the A stars (e.g. it is evidence of spots or pulsations) rather than being attributed to the spectral signatures of stellar companions. Variability is also apparent in the LSD profiles of KIC~5436432, KIC~6106152, KIC~7383872, and KIC~8692626 that is likely due to stellar companions based on the fact that (1) between two and three distinct components are visible and that (2) the components all exhibit large-scale radial velocity shifts. For each of the 44 stars in our sample, we computed additional LSD profiles using a line mask generated for a cool star with $T_{\rm eff}=4\,000\,{\rm K}$, $\log{g}=5.0\,{\rm (cgs)}$, and $\xi=2\,{\rm km\,s}^{-1}$. These profiles were used to identify additional (cooler) stellar components in KIC~7974841 and KIC~10815604. In Fig. \ref{fig:ex_lsd} we show several examples of the LSD profiles that were found to exhibit variability.

Radial velocities for all of the stellar components that were identified within the LSD profiles were derived by fitting a Gaussian function to each component thereby allowing the center of each line to be identified. For those cases in which (i) more than one component is visible in the associated LSD profiles and (ii) each component's core can be easily identified, multiple Gaussian functions were fit. Uncertainties in each radial velocity measurement were estimated by bootstrapping the residuals. This involved generating 100 bootstrapped data sets through random sampling of the residuals with replacement and adding these to the Gaussian fit; the fitting procedure was then carried out on each of the bootstrapped data sets and the uncertainties were taken to be 3 times the standard deviation of the resulting distribution. We note that for such a bootstrapping analysis, it is preferable to carry out a larger number of bootstrapped data sets (e.g. 500); however, we found that the distributions obtained using fewer bootstrapped data sets exhibit comparable standard deviations to those obtained using a larger number of bootstrapped data sets. We therefore used 100 data sets in order to reduce the computation time.

Comparing the $v_r$ values derived using the two methods (i.e. using spectral models and using LSD profiles), we find that the values and uncertainties ($\sigma_{v_r}$) are generally in agreement. The median $\sigma_{v_r}$ values derived using the spectral modelling and LSD methods is $2.7\,{\rm km\,s}^{-1}$, and $2.4\,{\rm km\,s}^{-1}$, respectively. All of the radial velocities derived from both the spectra and the LSD profiles are listed in the electronic version of this paper. The search for radial velocity variability and its interpretation is presented below in Sect. \ref{sect:binarity}.

\subsection{Chemical Abundances}\label{sect:abund}

As discussed in Sect. \ref{sect:spec_fit}, [M/H] values (metallicities) were derived during step one of the spectral modelling analysis. The metallicity roughly corresponds to the overall abundance of those elements heavier than He relative to the solar abundance table. A more detailed analysis is required to more accurately derive each individual element's chemical abundance, particularly in the case of chemically peculiar stars, which often exhibit significant abundance enhancements of only a small number of elements.

We derived chemical abundances for 41 of the 44 stars in our sample for between 1 and 12 elements per star depending largely on the S/N values associated with the averaged observations and the star's $v\sin{i}$ values (lower S/N values and higher $v\sin{i}$ values generally resulted in the abundances of fewer elements being constrained). For nearly all of the 41 stars, the abundances of Si and/or Fe-peak elements such as Fe and Cr were able to be derived primarily because of the prevalance of these lines within the $T_{\rm eff}$ range spanned by our sample. In a number of cases, the atmospheric abundances of rare earth elements such as Y, La, and Nd were able to be derived; this is particularly useful considering that Ap/Bp stars commonly exhibit significant enhancements in these particular abundances that can be several orders of magnitude higher than the abundances found in the Sun \citep[e.g.][]{preston1974,smith1996,ghazaryan2018}.

For the majority of the elements that were selected for abundance determination using the procedure described in Sect. \ref{sect:spec_fit}, only upper limits could be established. In these cases, the upper limits that were obtained were typically found to be very high \citep[i.e. above the chemical abundance excesses typically associated with CP stars; ][]{ghazaryan2018} and are therefore not reported here. For those elements for which both lower and upper limits could be derived, we report the weighted average. Generally, the abundance uncertainties yielded by the {\sc gssp} code are asymmetric; therefore, the weighted averages and their associated uncertainties were computed using the method described by \citet{barlow2004}.

\begin{figure}
	\centering
	\includegraphics[width=1.0\columnwidth]{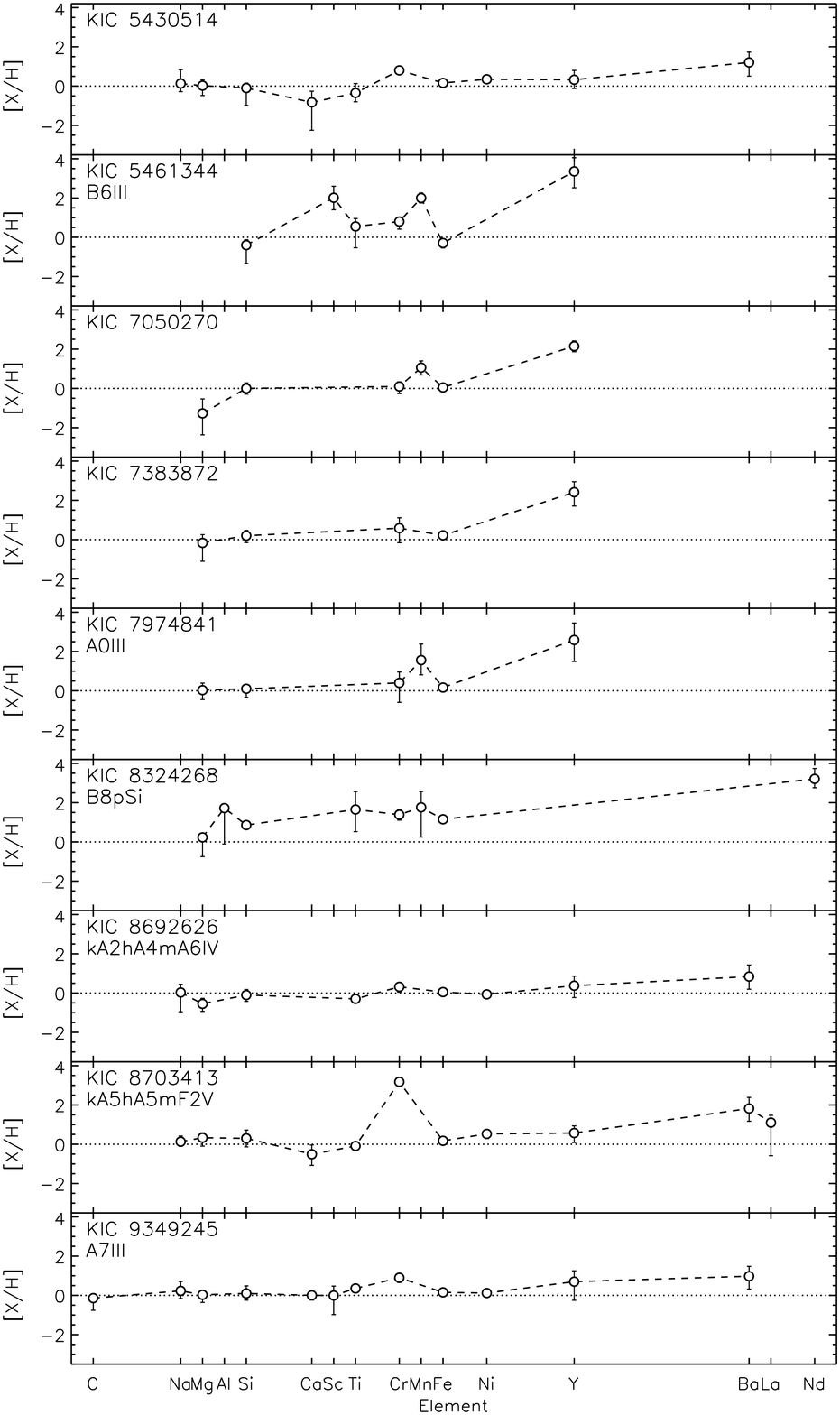}
	\caption{The derived chemical abundances, measured relative to the Sun's chemical abundances \citep{asplund2009}, of the stars with detected overabundancies of at least one element; those elements for which only upper bounds were derived are not shown.}
	\label{fig:abund01}
\end{figure}

We find that the majority of the 41 stars for which chemical abundances were derived have abundances that are approximately consistent with solar values or are slightly underabundant relative to solar values \citep{asplund2009}. However, 9 of the 41 stars have obvious overabundances of two or more elements; the derived abundances for these 9 stars are shown in Fig. \ref{fig:abund01}. The strongest overabundances are associated with KIC~8324268, which is identified in the literature as a Bp star; we observed strong overabundances of Si, Cr, Fe, and Nd with a high significance ($\gtrsim5\sigma$). Furthermore, this star exhibits line profile variability that was clearly detected in the LSD profiles, which is commonly associated with Bp stars.

Three of the 9 stars found to be overabundant in various elements are identified in the literature as Am stars (KIC~8692626, KIC~8703413, and KIC~9349245) were found to have enhanced abundances of various elements including Cr, Mn, Fe, Ni, Y, and Ba. Six of the 9 stars are not identified in the literature as being CP stars. KIC~5461344, KIC~7050270, and KIC~7974841 all have overabundances of Mn and exhibit relatively narrow spectral lines ($\lesssim30\,{\rm km\,s}^{-1}$) suggesting that they may belong to the HgMn class of CP stars. We note that KIC~5461344 also exhibits weak line profile variability, which is evident in the LSD profiles (see Fig. \ref{fig:ex_lsd}). Such a phenomenon has been previously found to be associated with a number of HgMn stars \citep[e.g.][]{ryabchikova1999,adelman2002,kochukhov2005}.

\subsection{SED fitting}\label{sect:SED_fit}

We carried out the synthetic energy distribution (SED) fitting analysis of available photometric observations and published distance estimates in order to derive each star's radius and luminosity. Photometric observations obtained using various filters are available for all of the 44 stars in our sample. This includes Johnson $B$ and $V$ measurements, Tycho $B_T$ and $V_T$ measurements \citep{esa1997}, 2MASS $J$, $H$, and $K_S$ measurements \citep{cohen2003}, and in several cases, Str{\" o}mgren $uvby$ measurements \citep{hauck1998} and/or Geneva $UB_1BB_2VV_1G$ measurements \citep{rufener1988}. Distances to 43 of the 44 stars have been published by \citet{bailer-jones2018} based on \emph{Gaia} DR2 parallax measurements; no distance estimate could be found for one of the 44 stars, KIC~5371784. The uncertainties associated with the distances are relatively low: the maximum uncertainty is $\approx7$~per~cent while the median is $\approx2$~per~cent.

The SED fitting analysis was carried out using the method described by \citet{sikora2019b} with models published by \citet{castelli2003}. In our application of this analysis, we fixed $T_{\rm eff}$, $\log{g}$, and [M/H] at the values derived from the spectroscopic observations (see Sect. \ref{sect:spec_fit}) while allowing the radius ($R$) to vary. We note that the grid of model SEDs have a range in [M/H] from $-4.0$ to $0.5$. For KIC~8324268, we derived a value of ${\rm [M/H]}=0.8$, which falls outside of this range; in this case we used the ${\rm [M/H]}=0.5$ SED models. Extinction values in the $V$ band ($A_V$) are reported by \citet{mathur2017} for 42 of the 44 stars in our sample and by \citep{brown2011} for the remaining 2 stars. The corresponding E$(B-V)$ values reported by these studies, which range from E$(B-V)=0.006\,{\rm mag}$ to $0.113\,{\rm mag}$, depend in part on the estimated $T_{\rm eff}$, $\log{g}$, and [M/H] values; therefore, inaccuracies in the fundamental stellar parameters will propagate to the E$(B-V)$ values. Given that we have derived estimates of $T_{\rm eff}$, $\log{g}$, and [M/H] spectroscopically, we opted to include E$(B-V)$ as a free parameter (where we impose the constraint that E$(B-V)\geq0$) in addition to $R$. The dereddening was carried out using the method described by \citet{cardelli1989}. The luminosities were then calculated using the Stefan-Boltzmann relation with the $T_{\rm eff}$ values and the radii derived from this analysis.

The inferred $R$ and $L$ values may be impacted, to some extent, by binarity: if the primary star has a stellar companion, its true $R$ and $L$ values will be lower than the derived values due to the contribution from the companion. Assuming that (i) any companion has a lower radius and effective temperature than that of the primary component and that (ii) the $T_{\rm eff}$ values derived from the spectra are accurate (Sect. \ref{sect:spec_fit}), the primary star's true $R$ and $L$ values may be overestimated by factors of $\leq\sqrt{2}$ and $\leq2$, respectively. In the 6 cases in which the spectral signatures of at least one dimmer companion was detected, their contributions to the total spectrum are significantly lower than those of the primary component; therefore, it is unlikely that the adopted $R$ and $L$ values are significantly affected by binarity.

For 42 of the 44 stars in our sample, revised $T_{\rm eff}$, $\log{L}$, and $R$ values have been published by \citet{mathur2017} and \citet{berger2018} based in part on the \emph{Gaia} DR2 catalog \citep{gaiacollaboration2018}. Comparing the $T_{\rm eff}$ values reported in these catalogs ($T_{\rm eff}^{\rm lit}$) with the values derived in this study ($T_{\rm eff}^{\rm spec}$, Sect. \ref{sect:spec_fit}), we find reasonable agreement where $T_{\rm eff}^{\rm spec}\lesssim9\,300\,{\rm K}$. Above $\approx9\,300\,{\rm K}$, the systematic discrepancy ($T_{\rm eff}^{\rm spec}-T_{\rm eff}^{\rm lit}$) increases roughly monotonically to $\sim2\,000{\rm K}$. This is perhaps unsurprising considering that (1) as noted in Sect. \ref{sect:sample}, the $T_{\rm eff}$ values reported in the KIC are likely unreliable above $9\,000\,{\rm K}$ \citep{brown2011} and (2) the $T_{\rm eff}$ values reported by \citet{mathur2017} for the 42 sample stars have simply been increased by $223\,{\rm K}$ relative to the KIC values. The radii derived in this study are largely discrepant with those values reported in the KIC. However, we find reasonable agreement between the radii reported by \citet{berger2018} and our values likely due to the fact that both studies rely on distance estimates derived from \emph{Gaia} DR2 parallax measurements \citep{bailer-jones2018}.

\subsection{Hertzsprung-Russell diagram}\label{sect:HRD}

The masses and ages of the stars in our sample were derived by comparing each star's Hertzsprung-Russell diagram (HRD) position as implied by its $T_{\rm eff}$ and $L$ values with published model evolutionary tracks. We used the grids published by \citet{mowlavi2012} and \citet{ekstrom2012}. The grid computed by \citet{mowlavi2012} has a higher mass and metallicity resolution but only extends from zero age main sequence masses of $0.5$ to $3.5\,M_\odot$. The majority of our stars (approximately 75~per~cent) have masses less than $3.5\,M_\odot$; in these cases, we used the high resolution grid to compute the masses and ages. For the stars with $T_{\rm eff}$ and $L$ values falling outside of the ranges associated with the evolutionary tracks computed by \citet{mowlavi2012} (and thus, with masses $\gtrsim3.5\,M_\odot$), we used the lower resolution grid computed by \citet{ekstrom2012}. In both cases, only the non-rotating models were used.

\begin{figure}
	\centering
	\includegraphics[width=1.0\columnwidth]{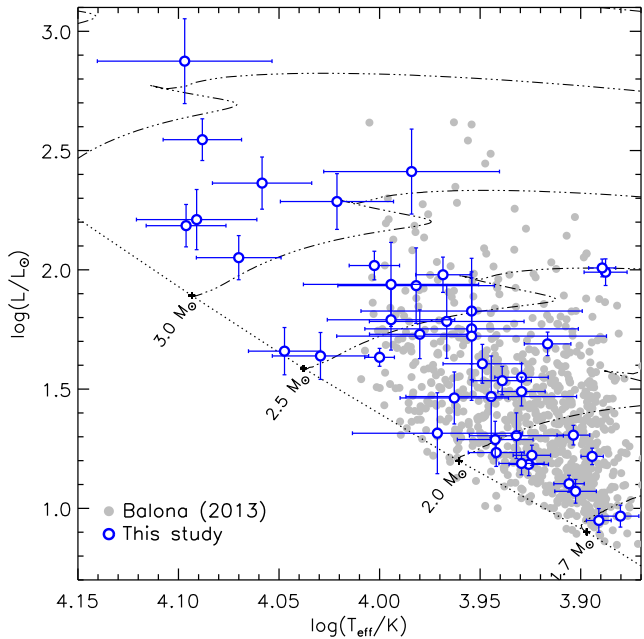}
	\caption[HRD associated with the stars composing the spectroscopic survey.]{HRD associated with the stars composing the spectroscopic survey (open 
	blue) using $T_{\rm eff}$ and $\log{L}$ values derived in this study. Filled grey 
	symbols correspond to the stars identified by \citet{balona2013} as 
	exhibiting variability that is consistent with rotational modulation; the 
	$T_{\rm eff}$ and $\log{L}$ values of these stars are taken either from the catalog compiled by \citet{mathur2017} and \citet{berger2018} (816 stars) or from the KIC \citep{brown2011} (59 stars).}
	\label{fig:HRD}
\end{figure}

The method by which the masses, ages, and their associated uncertainties were derived for the stars in our sample is described by \citet{sikora2019}. In Fig. \ref{fig:HRD}, we show the HRD along with the solar-metallicity, non-rotating models computed by \citet{ekstrom2012} for reference. The derived masses and ages are listed in the electronic version of this paper.

\subsection{Balmer line emission}\label{sect:balmer_emission}

\begin{figure}
	\centering
	\includegraphics[width=0.7\columnwidth]{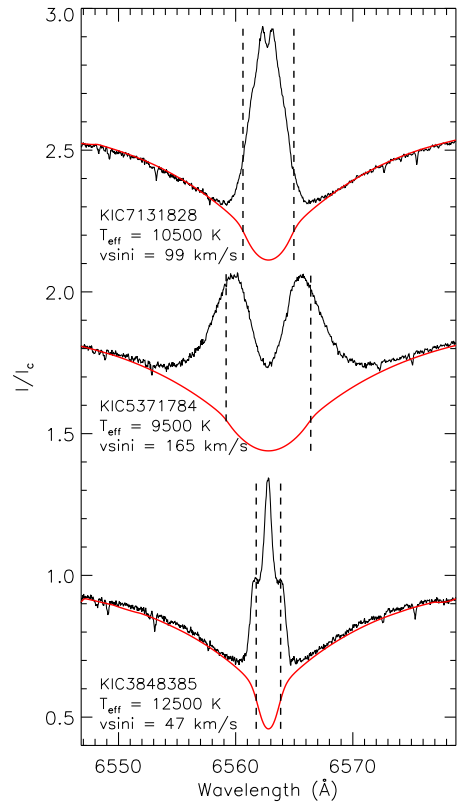}
	\caption{Observed (black) and synthetic (red) H$\alpha$ profiles for the three stars in our sample exhibiting strong, apparently non-variable emission. The vertical dashed lines indicate the extent of the Doppler-broadened core; in all three cases, the peak of the emission falls within these limits.}
	\label{fig:Halpha_emission}
\end{figure}

As noted in Sect. \ref{sect:spec_fit}, three of the stars in our sample, KIC~3848385, KIC~5371784, and KIC~7131828, were found to exhibit strong, apparently non-variable Balmer line emission. In Fig. \ref{fig:Halpha_emission}, we show the averaged H$\alpha$ line profiles for these three stars compared with model photospheric profiles generated using each star's derived fundamental parameters. It is evident that, in each of the three cases, the peak (or peaks, in the case of KIC~5371784 and KIC~7131828, which are characterized by two symmetric peaks on either side of the Balmer line's core) of the emission lies within the extent of the Doppler-broadened line core.

\begin{figure*}
	\centering
	\begin{minipage}{0.7\columnwidth}
		\centering
		\includegraphics[width=1.0\columnwidth]{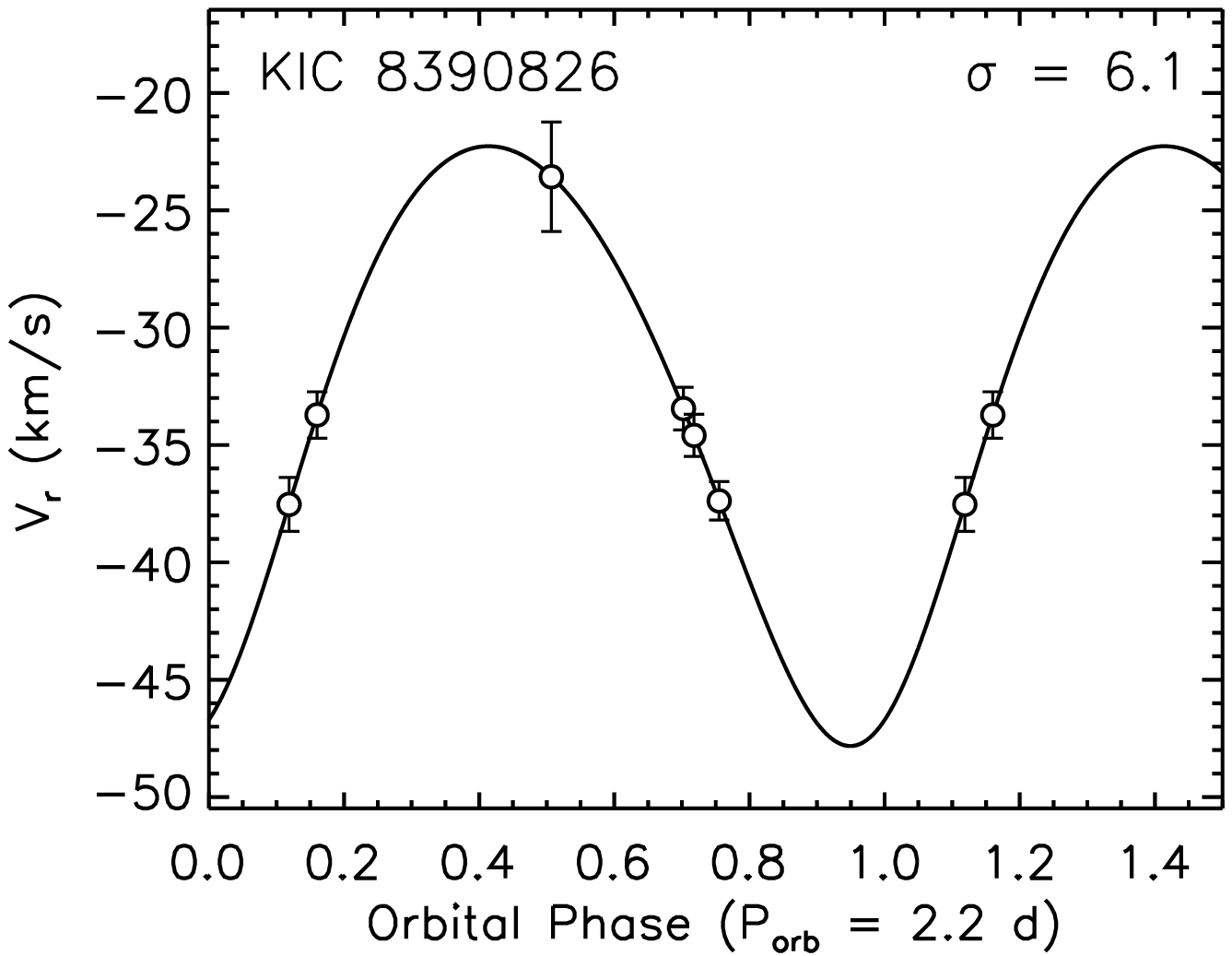}
	\end{minipage}\hspace{-0.2cm}
	\begin{minipage}{0.7\columnwidth}
		\centering
		\includegraphics[width=1.0\columnwidth]{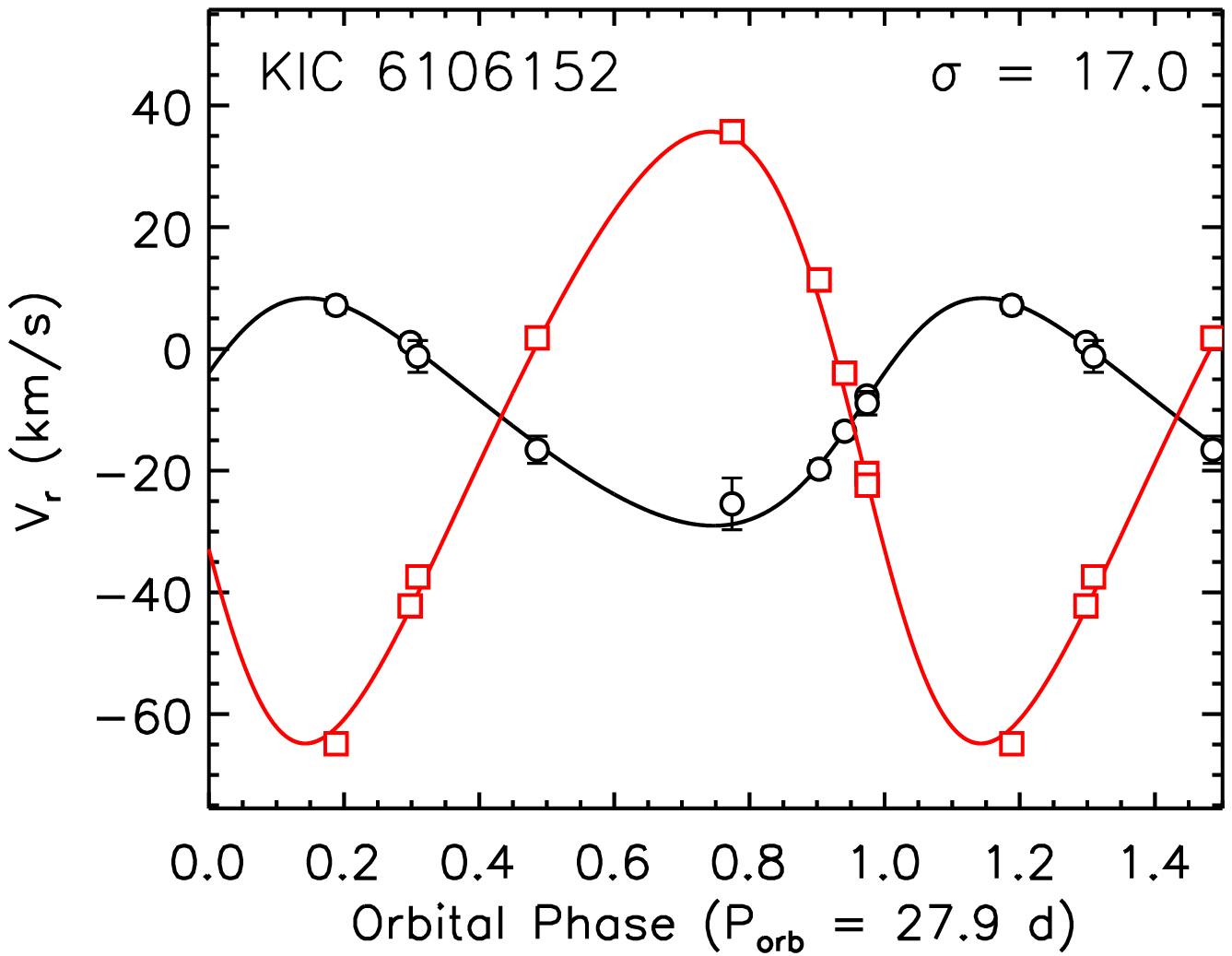}
	\end{minipage}
	\begin{minipage}{0.7\columnwidth}
		\centering
		\includegraphics[width=1.0\columnwidth]{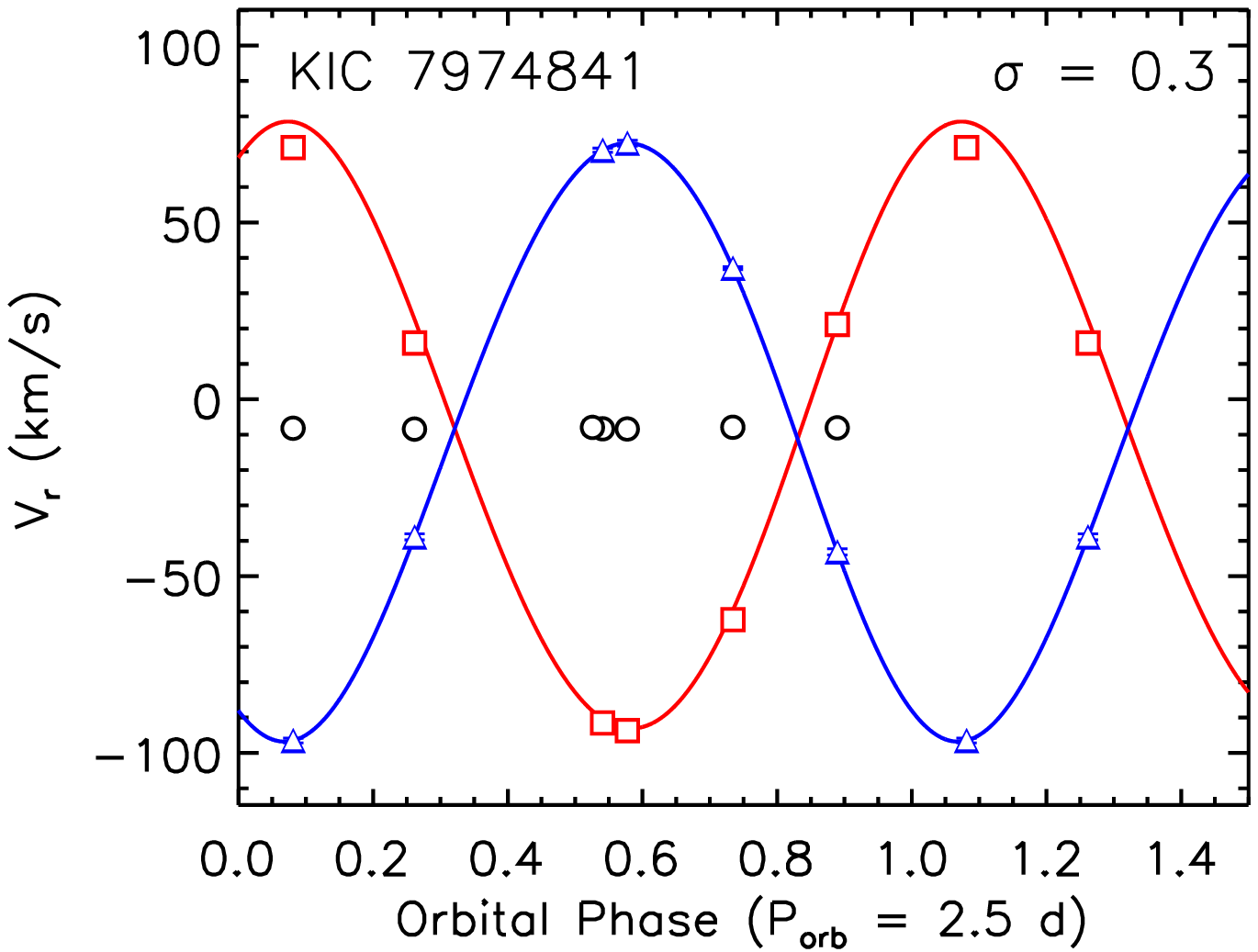}
	\end{minipage}
	\caption{Examples of the $v_r$ measurements phased by the best-fitting orbital periods associated with the primary component (KIC~8390826), the primary and secondary components (KIC~6106152), and the secondary and tertiary components (KIC~7974841). Black circles, red squares, and blue circles correspond to the primary, secondary, and tertiary components, respectively. The significance of $v_r$ variability ($\sigma$) of the primary components are listed in the top right corner of each panel.}
	\label{fig:ex_rv_curves}
\end{figure*}

Balmer line profiles similar to those shown in Fig. \ref{fig:Halpha_emission} are known to be associated with classical Be stars -- rapidly rotating MS B-type stars that host hot, gaseous Keplerian decretion disks \citep{slettebak1988,porter2003,rivinius2013}. The diversity of Be star Balmer line emission profiles is understood to be related principally to the inclination angle ($i$) of the star's rotation axis as proposed by \citet{struve1945} \citep[e.g. Fig. 3 of][]{slettebak1979} and demonstrated in Fig. 6 of \citet{sigut2015}; based on their Balmer line emission profiles, we would expect KIC~3848385 to have the lowest $i$, KIC~5371784 to have the highest $i$, and for KIC~7131828 to have an intermediate $i$ value. This can be tested using the derived $v\sin{i}$ values, $R$ values, along with the $P_{\rm rot}$ values inferred from the \emph{Kepler} light curves (Sect. \ref{sect:phot_var}). No distance estimate is available for KIC~5371784 and thus, its stellar radius could not be derived. For KIC~3848385, we find $i=11\pm3\,{\rm degrees}$ while for KIC~7131828, we find $i=20\pm5\,{\rm degrees}$; therefore the manner in which these two $i$ values differ is consistent with the geometrical model proposed by \citet{struve1945}. We also note that Be star Balmer line emission is known to be variable over timescales as long as $\sim$~decades \citep{rivinius2013}; considering that the three stars in our sample were observed over timescales $\sim1\,{\rm day}-9\,{\rm months}$, it is not surprising that we do not detect any variability in the Balmer line emission. 

\section{Binarity}\label{sect:binarity}

As noted in Sect. \ref{sect:radial_velocities}, we detected at least two stellar components associated with six stars in our sample. Additional multi-star systems were also inferred based on radial velocity variations of the primary star.

We searched the derived radial velocities for variability by first calculating generalized Lomb-Scargle (GLS) periodograms as described by \citet{zechmeister2009}. These periodograms were calculated using an oversampling factor of 10 and a maximum frequency of $1/\Delta t_{\rm min}$. The frequency exhibiting the maximum power was identified and used as the initial guess for the best-fitting orbital frequency ($f_{\rm orb}=1/P_{\rm orb}$). We evaluated the significance of any $v_r$ variability by comparing the sinusoidal fit associated with the best-fitting frequency yielded by the GLS periodogram to that of a constant (i.e. straight line) fit. The $\chi^2$ difference ($\Delta\chi^2$) between the fits was then computed and those cases in which $\Delta\chi^2$ corresponds to values $>5\sigma$ were considered to be significant and thus, members of multi-star systems. Statistically significant $v_r$ variability was detected for 6 of the stars in our sample for which no companions could be spectroscopically identified. Including those stars with spectroscopically identified companions, we conclude that at least 11 of the 44 stars ($25\pm11$~per~cent) in our sample are members of spectroscopic multi-star systems. This is consistent with the $\approx30-45$~per~cent incidence rate of spectroscopic binaries among intermediate-mass field stars \citep[e.g.][]{duchene2013}.

The GLS periodogram normalized power associated with false alarm probabilities (FAPs) of 1~per~cent for each set of radial velocities were estimated using a Monte Carlo simulation \citep[e.g.][]{cumming2004a}. In general, a large number ($\gtrsim10$) of possible orbital periods are associated with each of the 11 detected multi-star systems. This is likely related to the small number of observations obtained for each target. Nonetheless, we refined the best-fitting orbital periods inferred from the periodograms and derived orbital parameters to fit the $v_r$ curves. This was done by defining a grid of orbital periods centered on the estimated $P_{\rm orb}$ value with a grid resolution ranging from $10^{-2}$ to $10^{-6}\,{\rm d}$. Keplerian $v_r$ curves were then fit to each set of $v_r$ measurements using the {\sc mpfitfun} Levenberg-Marquardt algorithm implemented in {\sc idl}. The fitting carried out for each $P_{\rm orb}$ grid value involved four free parameters: the semi-amplitude ($K$), the systemic velocity ($v_0$), the longitude of periastron ($\omega$), and the eccentricity ($e$). The solution yielding the minimum $\chi^2$ value was then adopted. In the case of KIC~5436432, the best-fitting orbital solution yielded $e\approx0.7$, which led to large uncertainties in $v_0$ and $K_1$. The small number of available $v_r$ measurements do not allow for significant constraints to be placed on KIC~5436432's eccentricity; therefore, we enforced $e<0.6$ resulting in uncertainties that are comparable to the other identified SB1 systems.

\begin{table*}
	\caption{Derived orbital parameters for those systems found exhibiting statistically significant $v_r$ variability. Note that for each system, more than one possible orbital period was identified; additional measurements are necessary to better constrain the orbtial paramters. The first block lists the parameters derived only from the primary component's $v_r$ measurements, the second block (KIC~6106152) lists the parameters derived using both the primary and secondary components, and the third block (KIC~7974841) lists the parameters derived using the secondary and tertiary components where no $v_r$ variability associated with the primary component was detected.}
	\label{tbl:orb_param}
	\begin{center}
	\begin{tabular}{@{\extracolsep{\fill}}l c c c c c c c c@{\extracolsep{\fill}}}
		\noalign{\vskip-0.1cm}
		\cline{1-7}
		\cline{1-7}
		\noalign{\vskip0.5mm}
		KIC ID  &  $P_{\rm orb}\,({\rm d})$ & $v_0\,({\rm km\,s}^{-1})$ & $e$ & $K_1\,({\rm km\,s}^{-1})$ & $f(M)$                   & \multicolumn{1}{r}{$M_2^{\rm min}\,(M_\odot)$} & & \\
		        &                           &                           &     &                           & $(\times10^{-3}M_\odot)$ &                                                & & \\
		\noalign{\vskip0.5mm}
		\cline{1-7}
		\noalign{\vskip0.5mm}
		4567097      &         $6.6589(1)$ &       $-23.4\pm0.6$ &              $<0.3$ &         $4.7\pm0.7$ &              $<0.3$ &     \multicolumn{1}{r}{$0.10$} & & \\
		5436432      &         $57.595(2)$ &       $-35.5\pm0.5$ &              $>0.6$ &            $33\pm2$ &          $107\pm17$ &     \multicolumn{1}{r}{$0.93$} & & \\
		7383872      &       $1.017710(2)$ &       $-18.4\pm0.3$ &         $0.3\pm0.2$ &             $5\pm1$ &             $<0.03$ &     \multicolumn{1}{r}{$0.05$} & & \\
		8390826      &        $2.15721(2)$ &           $-33\pm1$ &              $<0.3$ &            $13\pm3$ &                $<1$ &     \multicolumn{1}{r}{$0.13$} & & \\
		8692626      &         $236.29(6)$ &           $-16\pm1$ &              $<0.8$ &             $3\pm1$ &                $<9$ &     \multicolumn{1}{r}{$0.16$} & & \\
		8703413      &         $6.5267(3)$ &       $-19.7\pm0.7$ &             $<0.03$ &      $25.65\pm0.08$ &             $<1100$ &     \multicolumn{1}{r}{$0.39$} & & \\
		10724634     &         $7.1183(2)$ &       $-16.7\pm0.4$ &              $<0.3$ &         $4.1\pm0.5$ &              $<0.3$ &     \multicolumn{1}{r}{$0.07$} & & \\
		11189959     &          $203.8(1)$ &           $-14\pm1$ &              $<0.3$ &             $9\pm2$ &               $<80$ &     \multicolumn{1}{r}{$0.44$} & & \\
		\cline{1-7} \\
		KIC ID  &  $P_{\rm orb}\,({\rm d})$ & $v_0\,({\rm km\,s}^{-1})$ & $e$ & $K_1\,({\rm km\,s}^{-1})$ & $K_2\,({\rm km\,s}^{-1})$ & $q$ & \multicolumn{1}{r}{$M_2\,(M_\odot)$} & \\
		\noalign{\vskip0.5mm}
		\cline{1-8}
		\noalign{\vskip0.5mm}
		6106152      &         $27.853(2)$ &           $-11\pm1$ &       $0.17\pm0.07$ &            $19\pm1$ &        $50.3\pm0.2$ &       $0.37\pm0.03$ &  \multicolumn{1}{r}{$0.79\pm0.08$} & \\
		\cline{1-8} \\
		KIC ID  &  $P_{\rm orb}\,({\rm d})$ & $v_0\,({\rm km\,s}^{-1})$ & $e$ & $K_2\,({\rm km\,s}^{-1})$ & $K_3\,({\rm km\,s}^{-1})$ & $q$ & $M_2^{\rm min}\,(M_\odot)$ & $M_3^{\rm min}\,(M_\odot)$ \\
		\noalign{\vskip0.5mm}
		\hline	
		\noalign{\vskip0.5mm}
		7974841      &       $2.510131(7)$ &        $-9.7\pm0.3$ &     $0.032\pm0.004$ &        $85.8\pm0.4$ &        $84.6\pm0.4$ &       $1.01\pm0.06$ &              $0.62$ &              $0.63$ \\
		\noalign{\vskip0.5mm}
		\hline \\
	\end{tabular}
	\end{center}
\end{table*}

We derived mass functions ($f[M]$) and minimum companion masses ($M_B^{\rm min}$) for the six targets for which only the primary component's radial velocities could be measured. In the cases of KIC~5436432 and KIC~7383872, the three identified components all exhibit statistically significant variability, however, no consistent orbital period associated with the components could be derived. We therefore only fit the primary component $v_r$ curves and report the derived $f(M)$ and $M_B^{\rm min}$ values. Additional observations are required in order to constrain the orbital configuration. For KIC~6106152, the GLS periodograms associated with the primary and secondary $v_r$ measurements both exhibit a maximum power at $\approx27.9\,{\rm d}$; a refined $P_{\rm orb}$ was derived along with the mass ratio ($q=M_B/M_A$). Neither the primary or the secondary components of KIC~10815604 were found to exhibit statistically significant variability, suggesting that $P_{\rm orb}\gg1\,{\rm yr}$.

For KIC~7974841, the secondary (B) and tertiary (C) components appear to exhibit similar amplitudes and orbital periods, however, they have similar spectroscopic signatures and could not be unambiguously distinguished in the LSD profiles. We remedied this by using the method described by \citet{hareter2008} which leverages the fact that, assuming that the two components are gravitationally bound, the quantity $|v_{r,B}-v_{r,C}|$ will vary with half the orbital period. We generated a GLS periodogram using the $|v_{r,B}-v_{r,C}|$ values, identified the best-fitting period, and labeled the B and C components based on the phased $v_{r,B}$ and $v_{r,C}$ curves. An orbital solution was then derived along with minimum masses of the two components. No variability was detected for the primary component, suggesting that the gravitationally bound secondary and tertiary components orbit the primary component with a period $\gg1\,{\rm yr}$.

In Fig. \ref{fig:ex_rv_curves} we show several examples of the phased $v_r$ measurements. The derived orbital periods, $v_r$ semi-amplitudes, systemic velocities, eccentricities, and secondary and tertiary component mass constraints are listed in Table~\ref{tbl:orb_param}.

\section{Photometric variability}\label{sect:phot_var}

The 44 A- and B-type stars in our sample have been identified by \citet{balona2013} as exhibiting variability that is consistent with rotational modulation. We analyzed the available \emph{Kepler} light curves in order to (i) verify the presence of such variability and (ii) to derive high-precision estimates of what are believed to be these stars' rotation periods.

We searched for statistically significant variability by first calculating Lomb-Scargle (LS) periodograms \citep{lomb1976,scargle1982,Press2007} from the post-processed \emph{Kepler} light curves (we used LS periodograms rather than the weighted GLS periodograms that were calculated for the $v_r$ measurements due to the prohibitively long computation times required to generate GLS periodograms from the light curves, which typically consist of $\gtrsim10^4$ measurements). The iterative pre-whitening procedure described by \citet{degroote2009a} was then carried out. This involved selecting the highest amplitude signal that is apparent in the LS periodogram, fitting a sinusoidal model given by
\begin{equation}\label{prewhiten_eqn}
	\Delta Kp(t)=c+\sum_{j=1}^{n_f}A_j\sin{[2\pi(f_jt-\phi_j)]},
\end{equation}
to the original light curve where the $j$ indices are associated with each of the $n_f$ extracted frequencies, and then calculating a new LS periodogram from the resulting residuals. Each parameter in Eqn. \ref{prewhiten_eqn} ($c$, $A_j$, and $\phi_j$) is re-fit for each new frequency that is extracted. This process may be carried out indefinitely leading to the extraction of both real astrophysical signals in addition to apparent signals that are introduced by noise. In order to avoid this, we adopted the stopping criterion proposed by \citet{vanreeth2015a} in which the $A_j$ values are compared with their associated amplitudes obtained from the LS periodogram. If these two values differ by more than $50$ per cent, the pre-whitening procedure is terminated. We also imposed an additional constraint such that a maximum of 100 frequencies would be extracted, which reduced the computational time required for the analysis.

\begin{figure}
	\centering
	\includegraphics[width=1.0\columnwidth]{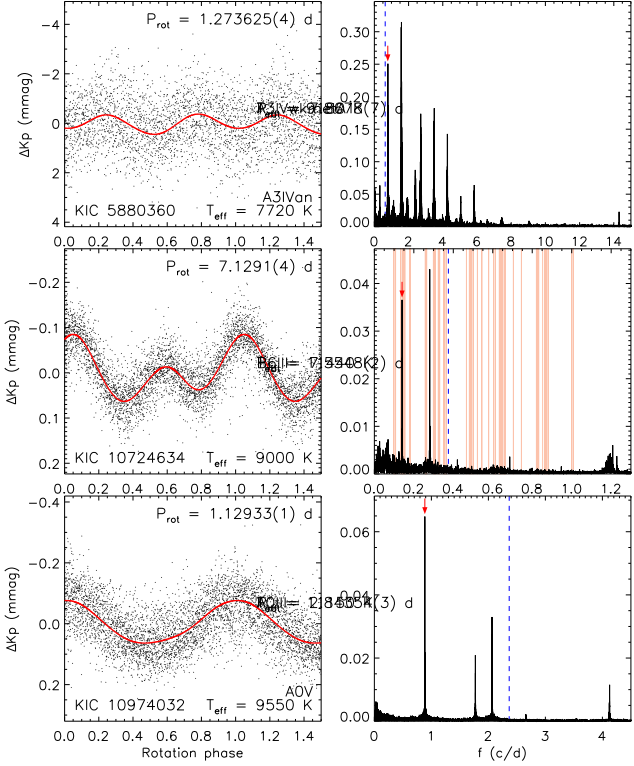}
	\caption{Examples of \emph{Kepler} light curves phased by the adopted rotation periods (left panels) that were found to exhibit multiple independent, low-frequency signals in their LS periodograms (right panels). The red curve corresponds to the best-fitting sinusoidal model (Eqn. \ref{prewhiten_eqn}). The adopted $P_{\rm rot}$ is indicated in the periodograms by a red arrow. The vertical dashed blue line indicates the minimum frequency that can be attributed to the star's rotation based on the derived $v\sin{i}$ values (i.e. rotation frequencies must be $\geq v\sin{i}/2\pi R$). The pink shaded regions that appear in some of the periodograms shown in here and in the online version correspond to orbital frequencies with FAPs $<1$~per~cent based on the $v_r$ measurements that exhibit statistically significant variability. Note that the light curves have been downsampled by a factor of 20 to reduce the figure file size.}
	\label{fig:ex_LC1}
\end{figure}

Many of the periodograms generated from the light curves exhibit relatively large amplitude peaks at $f\lesssim0.1\,{\rm c\,d^{-1}}$ that increase in amplitude with decreasing $f$; these signals are likely due to systematic instrumental effects \citep[e.g. quarterly gaps for data downloads, focus drift, quarterly rotations of the spacecraft,][]{balona2011a,murphy2012} and were removed from the lists of extracted frequencies. In several cases, the LS periodograms were found to contain regions of densely distributed, high-amplitude frequencies; in these cases, we removed all of the frequencies except for the one exhibiting the highest amplitude. The S/Ns of the extracted frequencies were estimated by assuming that the final LS periodogram calculated from the residuals of the final sinusoidal model (Eqn. \ref{prewhiten_eqn}) consists predominantly of noise (i.e. it is assumed that no real, high-amplitude signals remain in the final residuals). We then average the periodogram over frequency bins $0.1\,{\rm c\,d}^{-1}$ in width and divide each extracted frequency's amplitude by the local amplitude of the binned LS periodogram. For most of the light curves, the adopted pre-whitening frequency extraction analysis yielded multiple independent, low-frequency ($0.1\lesssim f\lesssim10\,{\rm c\,d^{-1}}$) signals, which may be attributed to pulsations, rotationally variable companion stars, the orbital period of those stars with detected companions, or a combination of these phenomena. Several examples of these light curves and their associated periodograms are shown in Fig. \ref{fig:ex_LC1}.

\begin{figure}
	\centering
	\includegraphics[width=1.0\columnwidth]{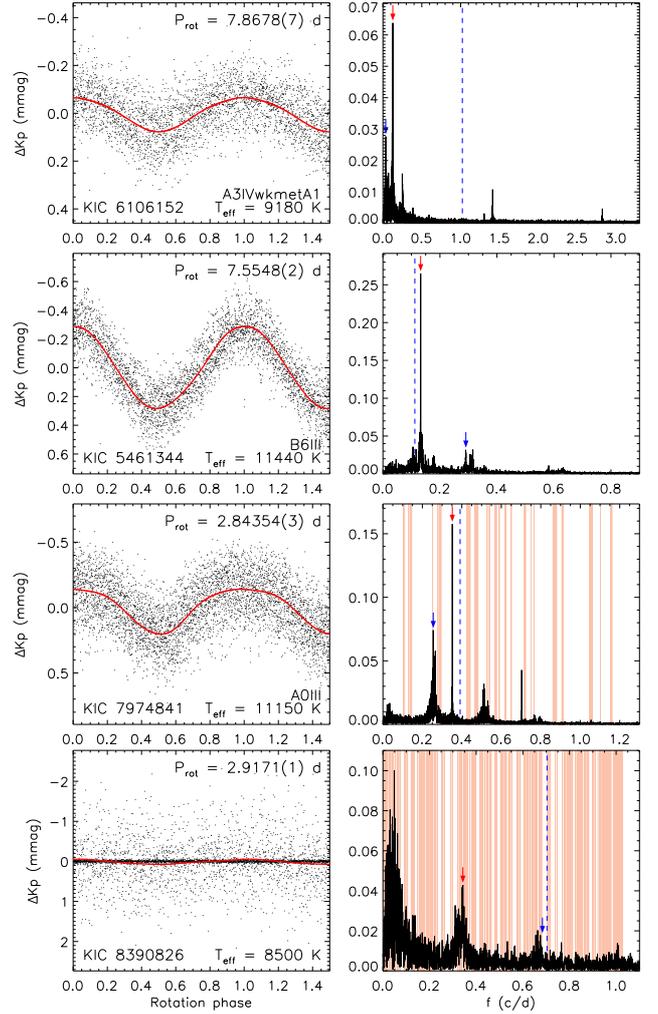}
	\caption{\emph{Kepler} light curves phased by the adopted rotation periods ($P_{\rm rot}$) (left panels) that differ significantly from those reported by \citet{balona2013} ($P_{\rm rot,B13}$) along with their associated LS periodograms (right panels). The red curve corresponds to the best-fitting sinusoidal model (Eqn. \ref{prewhiten_eqn}). The value of $1/P_{\rm rot}$ is indicated in the periodograms by a red arrow; a blue arrow indicates $1/P_{\rm rot,B13}$. The vertical dashed blue lines and pink shaded regions are defined in Fig. \ref{fig:ex_LC1}. Note that the light curves have been downsampled by a factor of 20 to reduce the figure file size.}
	\label{fig:ex_LC2}
\end{figure}

In general, the rotation periods derived in this study ($P_{\rm rot}$) are approximately consistent with those reported by \citet{balona2013} ($P_{\rm rot,B13}$); however, in 4 cases (KIC~6106152, KIC~5461344, KIC~7974841, and KIC~8390826), the adopted $P_{\rm rot}$ differs significantly from $P_{\rm rot,B13}$ (see Fig. \ref{fig:ex_LC2}). In these cases, the differences are related to the identification of the rotation frequencies in the periodograms: for those cases with multiple independent peaks, we selected the peak with the highest amplitude while \citet{balona2013} selected a lower amplitude peak.

The frequencies and amplitudes of the peaks that are presumed to be associated with each star's rotation were refined using the analysis described by \citet{sikora2019b} in which the light curves are fit using a sinusoidal function consisting of the rotation frequency and its first four harmonics. The rotation periods and the photometric amplitudes associated with these periods are listed in Table \ref{tbl:phot} along with the periods reported by \citet{balona2013}. All of the phased light curves and their associated LS periodograms are included in the online version of this paper.

\begin{table}
	\caption{Rotation periods and \emph{Kepler} photometric amplitudes associated with the 44 stars in our sample. Columns 1 to 4 list the KIC identifiers, rotation periods derived in this study, rotation periods reported by \citet{balona2013}, and the maximum photometric amplitudes inherent to the (presumably) rotationally modulated variability.}
	\label{tbl:phot}
	\begin{center}
	\begin{tabular}{@{\extracolsep{\fill}}l c c r@{\extracolsep{\fill}}}
		\noalign{\vskip-0.1cm}
		\hline
		\hline
		\noalign{\vskip0.5mm}
		KIC & $P_{\rm rot}$ & $P_{\rm rot,B13}$ & $\Delta Kp_{\rm max}$ \\
		ID  & $({\rm d})$   & $({\rm d})$       & (mmag)                \\
		(1) & (2)           & (3)               & (4)                   \\
		\noalign{\vskip0.5mm}
		\hline	
		\noalign{\vskip0.5mm}
1572201      &          2.36979(2) &       2.370 &   $0.0251(2)$ \\
2859567      &         0.491931(3) &       0.492 &   $0.0050(2)$ \\
3629496      &          0.42671(1) &       0.580 &   $3.9261(3)$ \\
3848385      &          1.69460(4) &       1.698 &   $0.0042(5)$ \\
4048716      &         0.586434(1) &       0.587 &   $0.0134(2)$ \\
4567097      &          2.78379(2) &       2.841 &   $0.0188(2)$ \\
4663468      &         0.438170(2) &       0.438 &   $0.0100(3)$ \\
4818496      &         0.614640(1) &       0.615 &   $0.0158(2)$ \\
4829781      &          2.94096(2) &       2.941 &   $0.0414(2)$ \\
4995049      &         0.734281(6) &       0.725 &   $0.0063(2)$ \\
5371784      &          1.00177(1) &       1.002 &   $0.1656(2)$ \\
5395418      &         0.598251(1) &       0.599 &   $0.0319(2)$ \\
5430514      &         1.355268(4) &       1.355 &   $0.0358(2)$ \\
5436432      &         3.270478(6) &       3.185 &   $0.1513(2)$ \\
5461344      &          7.55482(2) &       3.436 &   $0.2641(2)$ \\
5880360      &          1.27362(1) &       1.274 &   $0.3070(2)$ \\
6106152      &          7.86777(7) &      27.778 &   $0.0643(2)$ \\
6450107      &         0.652215(2) &       0.652 &   $0.0091(1)$ \\
7050270      &          10.9204(1) &      10.989 &   $0.0829(2)$ \\
7131828      &          0.62987(1) &       0.640 &   $0.0351(2)$ \\
7345479      &          0.43699(1) &       0.437 &   $0.0238(1)$ \\
7383872      &          6.84690(3) &       6.849 &   $0.1091(2)$ \\
7530366      &         0.749025(2) &       0.749 &   $0.0224(2)$ \\
7974841      &         2.843537(3) &       3.922 &   $0.1570(2)$ \\
8153795      &          0.45371(1) &       0.454 &   $0.0176(1)$ \\
8324268      &          2.00912(1) &       2.008 &  $12.3980(1)$ \\
8351193      &         0.571251(6) &       0.569 &   $0.0023(1)$ \\
8367661      &         0.624769(1) &       0.625 &   $0.0258(2)$ \\
8390826      &          2.91711(1) &       1.462 &   $0.0535(5)$ \\
8692626      &          1.64812(3) &       1.647 &   $0.0047(2)$ \\
8703413      &         6.526278(7) &       6.536 &   $0.2144(2)$ \\
9349245      &         0.942002(8) &       0.943 &   $0.0063(1)$ \\
9392839      &          4.47203(2) &      71.429 &   $0.0535(1)$ \\
9468475      &         0.646197(1) &       0.646 &   $0.0156(1)$ \\
9772586      &          0.51444(1) &       0.514 &   $0.0822(2)$ \\
10724634     &          7.12911(4) &       7.143 &   $0.0431(2)$ \\
10815604     &           8.8408(2) &       8.850 &   $0.0226(2)$ \\
10879812     &         0.699426(3) &       0.711 &   $0.0113(2)$ \\
10974032     &         1.129327(1) &       1.130 &   $0.0647(2)$ \\
11189959     &         0.822467(3) &       0.822 &   $0.0132(2)$ \\
11443271     &          0.80791(1) &       0.808 &   $0.0668(1)$ \\
11600717     &           2.2135(2) &       2.222 &   $0.0713(5)$ \\
12061741     &          2.94251(1) &       2.941 &   $0.0550(2)$ \\
12306265     &         0.886820(9) &       0.887 &   $0.0065(2)$ \\
		\noalign{\vskip0.5mm}
		\hline \\
	\end{tabular}
	\end{center}
\end{table}

\section{Testing the Rotational Modulation Hypothesis}\label{sect:test}

One of the primary goals of this study is to derive the rotational broadening parameters ($v\sin{i}$) for each of the stars in our sample, which can in principle be used to falsify the hypothesis that photometric variability is due to the star's rotation. This is carried out by comparing $v\sin{i}$ with the equatorial velocity ($v_{\rm eq}\equiv2\pi R/P_{\rm rot}$), which is derived by assuming that an identified photometric frequency ($f$) corresponds to the star's rotation frequency (i.e. $f=f_{\rm rot}=1/P_{\rm rot}$); if $f$ does correspond to $f_{\rm rot}$, then rigid rotation requires that $f\geq v\sin{i}/2\pi R$.

\begin{figure}
	\centering
	\includegraphics[width=1.0\columnwidth]{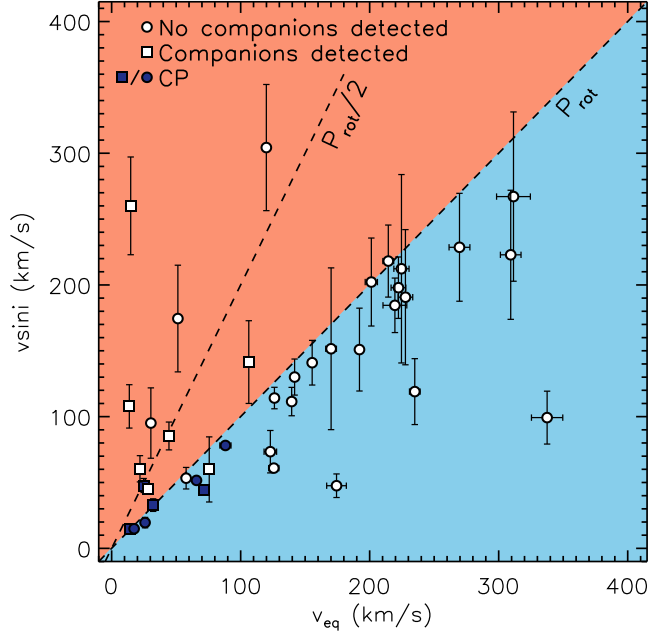}
	\caption{Comparison between the $v\sin{i}$ values derived from the spectroscopic observations and the $v_{\rm eq}$ values derived using each star's $R$ value and $P_{\rm rot}$ value inferred from the Kepler light curves. Points appearing in the top-left are inconsistent with rotational modulation while points in the lower-right are consistent with rotational modulation. Filled blue points correspond to those stars for which chemical overabundances (i.e. higher than solar) for at least one element were detected. The dashed lines correspond to $v_{\rm eq}$ derived using $P_{\rm rot}$ and $P_{\rm rot}/2$. Note that each point has an assigned uncertainty and that in certain cases, the uncertainties are smaller than the size of the symbols.}
	\label{fig:vsini_veq}
\end{figure}

In Fig. \ref{fig:vsini_veq}, we compare the $v\sin{i}$ and $v_{\rm eq}$ values for the 43 stars in our sample with derived $v_{\rm eq}$ values. As discussed previously, no distance estimate is available for KIC~5371784, which is required to derive $R$ and by extension, $v_{\rm eq}$. We find that 33 of the 43 stars ($77\pm25$~per~cent) exhibit $v\sin{i}<v_{\rm eq}$ within the estimated uncertainties and therefore, the identified photometric periods can plausibly be attributed to each star's rotation period. In Figures \ref{fig:ex_LC1} and \ref{fig:ex_LC2} as well as in those figures presented in the online version of this paper, the periodograms are shown with the minimum $f$ that can be attributed to the star's rotation based on the derived $v\sin{i}$ values (i.e. $f_{\rm min}=v\sin{i}/2\pi R$). It is clear that for 9 of the 10 stars that have $v\sin{i}>v_{\rm eq}$ based on the adopted $P_{\rm rot}$ and derived $R$ (KIC~5436432, KIC~6106152, KIC~7383872, KIC~9392839, KIC~10724634, KIC~10815604, KIC~10974032, KIC~11189959, and KIC~12061741), there are additional periodogram peaks with $f>f_{\rm min}$. Therefore, although the $P_{\rm rot}$ values derived here along with those reported by \citet{balona2013} for these 9 stars are not physically compatible with each star's rotation period, it is plausible that those higher-frequency peaks do correspond to rotation periods.

\begin{figure}
	\centering
	\includegraphics[width=0.75\columnwidth]{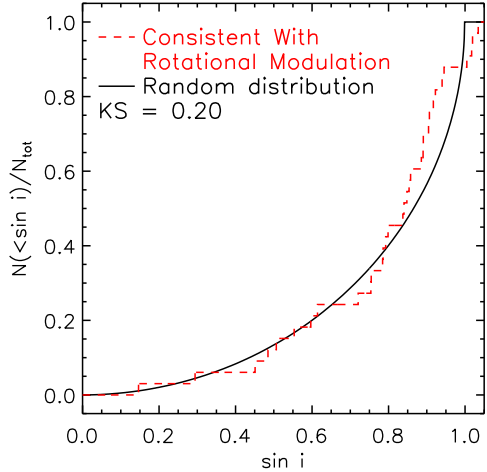}
	\caption[Cumulative distribution function of $\sin{i}$.]{Cumulative distribution function (CDF) of $\sin{i}$ (where $i$ are the inclination angles) for those stars in our sample consistent with $v\sin{i}<v_{\rm eq}$ (dashed red curve). The solid black curve shows the CDF expected from a distribution of stars with random inclination angles. The Kolmogorov-Smirnov (KS) test statistic comparing the two distributions is shown.}
	\label{fig:inc_cdf}
\end{figure}

We carried out an additional test of the rotational modulation hypothesis by deriving inclination angles ($\sin{i}$ values) for those 33 stars for which $v\sin{i}<v_{\rm eq}$ and comparing the resulting distribution with that expected from a sample of stars having randomly oriented rotation axes \citep[e.g.][]{jackson2010}. In Fig. \ref{fig:inc_cdf}, we compare the cumulative distribution functions (CDFs) of these two samples. We find no statistically significant difference between the two distributions based on the derived Kolmogorov-Smirnov (KS) test statistic of 0.20 and accompanying $p$-value of 0.12.

\section{Discussion}\label{sect:discussion}

The discovery by \citet{balona2013} that approximately 44 per cent of the MS A stars observed with \emph{Kepler} exhibit variability that is consistent with rotational modulation presents a challenge to our understanding of the physical constitution of these objects. It suggests that a significantly higher fraction of MS A stars may host inhomogeneous surfaces -- i.e. chemical or brightness spots -- than previously believed. Currently, the detection of chemical spots via both photometric rotational modulation and spectral line profile variability has only been reported for Ap/Bp stars, which account for $\sim10$ per cent of all MS A- and B-type stars, and for a small number of relatively rapidly rotating HgMn stars. Therefore, verifying whether similar phenomena are associated with 44 per cent of all MS A-type stars is necessary in order to address theoretical questions that have transpired over the past decade.

In this study, we have obtained a large number of high-resolution ESPaDOnS Stokes $I$ observations for 44 MS A and late-B type stars that were previously identified by \citet{balona2013} as being rotationally variable. There are two primary goals of this survey. First, to test the hypothesis that the photometric variability detected using \emph{Kepler} is in fact rotational modulation that is intrinsic to these objects. As described in Sect. \ref{sect:test}, this was carried out by determining whether $v\sin{i}<v_{\rm eq}$ where $v_{\rm eq}=2\pi R/P_{\rm phot}$ and $P_{\rm phot}$ is the period of photometric variability that is hypothesized to correspond to $P_{\rm rot}$ \citep[e.g.][]{balona2017,balona2019b}. Furthermore, our survey has been designed to search for radial velocity variability that may be indicative of low-mass ($\lesssim1\,M_\odot$) companions with convective envelopes, which commonly exhibit rotationally modulated light curves produced by star spots \citep[e.g.][]{mcquillan2012}. The detection of such companions provides a plausible, and arguably more probable, alternative explanation for the photometric variability. The second primary goal is to search for strong chemical peculiarities and/or strong spectral line profile variability that is commonly associated with Ap/Bp stars. The results of these two goals are broadly summarized in Table \ref{tbl:summary} and discussed in detail below.

\begin{table}
	\caption{Summary of the results of our survey. Column 2 indicates whether the photometric variability is consistent with rotational modulation (i.e. whether $v\sin{i}<v_{\rm eq}\equiv2\pi R/P_{\rm phot}$ within the uncertainties). Columns 3 to 5 indicate whether line profile variability, chemical peculiarites, or companions with masses potentially $\lesssim1\,M_\odot$ were detected. Column 6 indicates whether $P_{\rm phot}$ coincides with possible orbital periods for those stars with detected companions. Only 21 stars are listed; the remaining 23 stars in the sample exhibit $v\sin{i}<v_{\rm eq}$ and no detected line profile variability, chemical peculiarities, or companions.}
	\label{tbl:summary}
	\begin{center}
	\begin{tabular}{@{\extracolsep{\fill}}l c c c c r@{\extracolsep{\fill}}}
		\noalign{\vskip-0.1cm}
		\hline
		\hline
		\noalign{\vskip0.5mm}
		KIC ID & Rot  & Line & CP? & Low-$M$ & \multicolumn{1}{l}{$P_{\rm phot}$} \\
		       & var? & var? &     & comp?   & $\approx P_{\rm orb}$? \\
		(1)    & (2)  & (3)  & (4) & (5)     & (6) \\
		\noalign{\vskip0.5mm}
		\hline	
		\noalign{\vskip0.5mm}
5430514  & Yes &  No & Yes &  No & \\
5461344  & Yes & Yes & Yes &  No & \\
5880360  & Yes & Yes &  No &  No & \\
7050270  & Yes &  No & Yes &  No & \\
8324268  & Yes & Yes & Yes &  No & \\
9349245  & Yes &  No & Yes &  No & \\
\hline
4567097  & Yes &  No &  No & Yes &  No \\
7974841  & Yes &  No & Yes & Yes &  No \\
8692626  & Yes &  No & Yes & Yes &  No \\
8703413  & Yes &  No & Yes & Yes & Yes \\
\hline
5371784  &   ? &  No &  No &  No &     \\
5436432  &  No &  No &  No & Yes &  No \\
6106152  &  No &  No &  No & Yes &  No \\
7383872  &  No &  No & Yes & Yes &  No \\
8390826  &  No &  No &  No & Yes & Yes \\
9392839  &  No &  No &  No &  No &     \\
10724634 &  No &  No &  No & Yes & Yes \\
10815604 &  No &  No &  No & Yes &  No \\
10974032 &  No &  No &  No &  No &     \\
11189959 &  No &  No &  No & Yes &  No \\
12061741 &  No &  No &  No &  No &     \\
		\noalign{\vskip0.5mm}
		\hline \\
	\end{tabular}
	\end{center}
\end{table}

Our analysis indicates that 33 of the 43 stars in our sample for which $v_{\rm eq}$ could be inferred have $v\sin{i}<v_{\rm eq}$ within the estimated uncertainties (i.e. their periods and line widths are consistent with the rotational modulation hypothesis). Our search for radial velocity variability yielded detections of stellar companions associated with 11 of the 44 stars in our sample, which includes 4 of the 33 stars with $v\sin{i}<v_{\rm eq}$. The companions were all found to have minimum masses ranging from $\approx0.1$ to $1\,M_\odot$ (Table \ref{tbl:orb_param}). Moreover, the spectroscopic signatures of the companions were either not detected or were found to be relatively weak compared to that of the respective primary component indicating that they are dimmer and less massive. Therefore, it would not be surprising for these companion stars to host star spots that could be responsible for producing the observed photometric variability.

M dwarfs found in the \textit{Kepler} field for which rotational variability has been detected exhibit typical flux amplitudes ($\Delta F_{\rm M}/F_{\rm M}$) ranging from $\sim0.1-1$~per~cent \citep{mcquillan2014}. Assuming that the variability discussed in the present study is being produced by a mid-type M dwarf companion, we can roughly estimate the $\Delta F_{\rm M}/F_{\rm M}$ (i.e. the M dwarf's intrinsic variability) that is required to produced the observed $\Delta Kp$ amplitudes listed in Table \ref{tbl:phot}. Using the derived $T_{\rm eff}$ and $R$ values for the stars in our sample and the fact that mid-type M dwarfs have $T_{\rm eff}\sim3\,200\,{\rm K}$ and $R\sim0.3\,R_\odot$ \citep[e.g.][]{hardegree-ullman2019}, we find a median $\Delta F_{\rm M}/F_{\rm M}=4.6$~per~cent. For 14 stars in our sample ($\approx30$~per~cent), we find $\Delta F_{\rm M}/F_{\rm M}<1$~per~cent thereby supporting the notion that a non-negligible fraction of the stars in our sample exhibit variability that can plausibly be explained by the presence of M dwarf companions.

It is also plausible that the identified $P_{\rm phot}$ for those stars with identified companions may be associated with the orbital periods of these systems \citep[e.g. the systems may be ellipsoidal or eclipsing variables,][]{smalley2014}. As is shown in the periodograms (Figures \ref{fig:ex_LC1} and \ref{fig:ex_LC2} along with those included in the online version of this paper), 3 of the 11 stars with detected companions have $P_{\rm phot}$ values that coincide with possible orbital periods inferred from radial velocity measurements. We conclude that, for those 4 stars with $v\sin{i}<v_{\rm eq}$ and with detected companions, it is plausible that the photometric variability is not in fact intrinsic to the A/late-B type stars, however, we cannot definitively rule out the rotational modulation hypothesis.

Based on the observations obtained within this study, we conclude that the rotational modulation hypothesis remains a plausible explanation for the origin of the \emph{Kepler} photometric variability associated with the 29 of 43 MS A/late-B type stars in our sample ($67\pm23$~per~cent) that have $v\sin{i}<v_{\rm eq}$ and no detected low-mass companions. \citet{balona2013} reported that 875 out of the estimated $1\,974$ MS A type stars observed with \emph{Kepler} ($44\pm2$~per~cent) exhibit rotational modulation. Assuming that our sample of 43 stars is representative of the larger sample of 875 reportedly rotationally variable A stars, our results suggest that the true incidence rate of this phenomenon may be approximately $30\pm2$~per~cent. Although this fraction is significantly smaller than reported by \citet{balona2013}, it greatly exceeds the fraction of known magnetic A and late-B type stars that might be expected to exhibit rotational modulation.

Chemical abundances exceeding solar values in at least one element were detected for 9 stars of our sample suggesting that they are CP stars; of these, five have $v\sin{i}<v_{\rm eq}$ and no companions, 3 have $v\sin{i}>v_{\rm eq}$ and detected companions, and 1 has $v\sin{i}>v_{\rm eq}$ and no detected companions. Out of the 5 CP stars with $v\sin{i}<v_{\rm eq}$ and no companions, only 1 (KIC~8324268) is identified in the literature as an Ap/Bp star. Line profile variability was also detected for this star, along with the CP star KIC~5461344, which suggests that both are likely Ap/Bp stars and host strong magnetic fields with strengths $\gtrsim100\,{\rm G}$. The remaining 3 CP stars with $v\sin{i}<v_{\rm eq}$ and no detected companions (KIC~5430514, KIC~7050270, and KIC~9349245) all exhibit enhanced Cr, which is most commonly associated with Ap/Bp stars \citep[e.g.][]{ghazaryan2018}. In summary, we consider these 5 targets to be candidate magnetic Ap/Bp stars.

As shown in Table \ref{tbl:obs}, the sample contains 3 stars that have Am spectral types reported in the literature (KIC~8351193, KIC~8692626, and KIC~8703413). No abundances could be derived for KIC~8351193 as a result of its high $v\sin{i}$ value of $150\,{\rm km\,s}^{-1}$, which also suggests that it has been misclassified considering that Am stars typically have relatively low $v\sin{i}$ values \citep[e.g.][]{hauck1986,zorec2012}. The two other Am stars both have detected companions and $v\sin{i}$ values $\lesssim45\,{\rm km\,s}^{-1}$. In both of these cases the photometric variability can possibly be attributed to spots on the (presumably) low-mass companion. For KIC~8703413, $P_{\rm phot}$ coincides with a possible orbital period suggesting that the system may be an ellipsoidal variable \citep[e.g.][]{smalley2014}.

Removing the 5 Ap/Bp candidates along with those stars with detected companions leaves a total of 24 stars in our sample that (1) exhibit photometric variability that is consistent with rotational modulation (i.e. they have $v\sin{i}<v_{\rm eq}$) and (2) that are presumably not strongly magnetic based on the non-detection of either strong chemical peculiarities or spectral line profile variability. Currently, the origin of the photometric variability in these objects is unkown. It has been suggested that the variability may be related to ultra-weak magnetic fields having strengths $\lesssim1\,{\rm G}$ \citep{blazere2016,petit2017} similar to those theequation have been detected on a small number of objects such as Vega \citep{lignieres2009}. Assuming that the observed variability is related to a single mechanism \citep[e.g. magnetic fields generated in thin sub-surface convective layers,][]{cantiello2019}, we estimate that such a mechanism may be active in $\lesssim24$~per~cent of all MS A/late-B type stars where we have applied the same statistical argument that was used above.

\section{Conclusion}\label{sect:conclusion}

We have presented a detailed high-resolution spectroscopic study of 44 MS A/late-B type stars that reportedly exhibit rotationally modulated \emph{Kepler} light cuves \citep{balona2013}. For 14 of these stars, we found that either (1) the derived $v\sin{i}$ values are inconsistent with the photometric periods being attributed to the star's rotation period or (2) were found to have low-mass companions that may be producing the observed variability. We conclude that 29 of the 43 stars for which radii and $v\sin{i}$ values could be derived are likely consistent with the rotational modulation hypothesis since they have $v\sin{i}<v_{\rm eq}$ and no detected low-mass companions. This allowed us to revise the previously reported incidence rate of rotationally variable MS A type stars from $44\pm2$~per~cent \citep{balona2013} to an estimated $30\pm2$~per~cent.

Obtaining observations capable of detecting or ruling out the presence of chemical spots, temperature spots, or weak magnetic fields that may be associated with a large number of MS A type stars is an important next step for our understanding of the origin of these stars' photometric variability. Considering that the median $V$ magnitude for the stars in our sample is $8.4\,{\rm mag}$, detecting low-contrast spots or weak magnetic fields may require prohibitively long exposure times. Opportunities to search for such phenomena may become available in the near future as a result of the ongoing \emph{TESS} mission, which is allowing for the detection of photometric variability that may be attributed to rotational modulation of significantly brighter A and B-type stars \citep[e.g.][]{sikora2019b,balona2019a} than those located in the \emph{Kepler} field. 

The detection of either weak spots or weak magnetic fields that may be visible at the surfaces of such stars can be achieved by obtaining a series of extremely high S/N ($\gtrsim1\,000$) Stokes $I$ or $V$ spectra using high-resolution spectropolarimeters \citep[e.g.][]{lignieres2009,bohm2015} such as ESPaDOnS@CFHT or NARVAL@TBL. Suitable targets for such a search will need to be bright, have relatively short rotation periods (such that the rotation period can be densely sampled), and have intermediate $v\sin{i}$ values. Considering the high cost of detecting and characterizing these features, it may be most efficient to first obtain low- to moderate-S/N spectroscopic snapshot observations of bright \emph{TESS} targets that are found to exhibit rotational modulation in order to identify suitable targets. These targets may subsequently be followed up with the requisite high-S/N Stokes $I$ or $V$ observations.

\section*{Acknowledgments}

GAW acknowledges support in the form of a Discovery Grant from the Natural Science and 
Engineering Research Council (NSERC) of Canada.

\section*{Data availability}

The data underlying this article are available in the article and in its online supplementary material.

\bibliography{Kepler_AB}
\bibliographystyle{mn2e}

\end{document}